\documentclass[12pt]{article}
\usepackage{graphicx}
\usepackage{amstext}
\usepackage{caption}
\usepackage{etoolbox}
\usepackage{makeidx}
\include{epsf}
\usepackage[toc,page]{appendix}
\usepackage{amsfonts}
\usepackage{authblk} 
\usepackage{sectsty}
\usepackage{amsmath,amssymb,epsfig}
\usepackage{amscd}
\usepackage{amsthm}
\usepackage{braket}
\usepackage{amsmath}
\usepackage{xcolor}
\usepackage{amssymb}
\usepackage{mathrsfs}
\usepackage{setspace}
\usepackage{url}
\usepackage{dsfont}
\usepackage[applemac]{inputenc}
\usepackage[english]{babel}
\usepackage{enumitem} 
\usepackage{verbatim}
\usepackage{slashed}
\usepackage[]{latexsym}
\usepackage{mathtools}
\usepackage{subcaption}
\usepackage{ mathrsfs }
\usepackage{braket}
\usepackage{caption}
\usepackage{cite}
\usepackage{dsfont}
\usepackage{hyperref}
\hypersetup{
pdftitle={},%
pdfauthor={},%
pdfsubject={},%
pdfkeywords={},%
colorlinks=true,%
linkcolor=blue,%
citecolor=red,%
linktocpage=true,%
pageanchor=true
}

\newcommand{\be}{\begin{equation}}
\newcommand{\ee}{\end{equation}}
\def\beqa{\begin{eqnarray}}
\def\eeqa{\end{eqnarray}}
\def\bean{\begin{eqnarray*}}
\def\eean{\end{eqnarray*}}
\def\nn{ }

\newcommand{\R}{\mathbb{R}}
\newcommand{\C}{\mathbb{C}}

\newcommand{\dd}{{\mathrm{d}}}
\theoremstyle{definition}
\newtheorem{definition}{Definition}[section]

\newtheorem{lemma}{Lemma}[section]

\newcommand{\eqn}[1]{(\ref{#1})}
\newcommand{\del}{\partial}
\newcommand{\Tr}[1]{\:{\rm Tr}\,#1}

\renewenvironment{thebibliography}[1]
         {\section*{References}\frenchspacing\small
          \begin{list}{[\arabic{enumi}]}
         {\usecounter{enumi}\parsep=2pt\topsep 0pt
         \settowidth{\labelwidth}{[#1]}
         \leftmargin=\labelwidth\advance\leftmargin\labelsep
         \rightmargin=0pt\itemsep=1pt\sloppy}}{\end{list}}

 \numberwithin{equation}{section}

\title{\textbf{Poisson-Lie T-duality of WZW model  via current algebra deformation}\vspace{0.5cm}}
\date{}

\author[1,2]{Francesco Bascone}
\author[1]{Franco Pezzella}
\author[1,2]{Patrizia Vitale}
\affil[ ]{}
\affil[1]{\textit{\footnotesize INFN-Sezione di Napoli, Complesso Universitario di Monte S. Angelo Edificio 6, via Cintia, 80126 Napoli, Italy.}}
\affil[2]{\textit{\footnotesize Dipartimento di Fisica ``E. Pancini'', Universit\`a di Napoli Federico II, Complesso Universitario di Monte S. Angelo Edificio 6, via Cintia, 80126 Napoli, Italy.}}
\affil[ ]{}
\affil[ ]{\footnotesize e-mail: \texttt{francesco.bascone@na.infn.it, franco.pezzella@na.infn.it, patrizia.vitale@na.infn.it}}
\begin{document}
\maketitle
\begin{abstract}
\small  Poisson-Lie T-duality  of the Wess-Zumino-Witten (WZW) model having the group manifold of $SU(2)$ as target space is investigated. The whole construction relies on the deformation of the affine current algebra of the model, the semi-direct sum  $\mathfrak{su}(2)(\mathbb{R}) \, \dot{\oplus} \, \mathfrak{a}$, to the fully semisimple Kac-Moody algebra $\mathfrak{sl}(2,\mathbb{C})(\mathbb{R})$. 
A two-parameter family of models with $SL(2,\C)$ as target phase space is obtained so that  Poisson-Lie T-duality is realised as an  $O(3,3)$ rotation in the phase space. The dual  family  shares the same phase space but its configuration space is $SB(2,\mathbb{C})$, the Poisson-Lie dual of the group $SU(2)$.  A parent action with doubled degrees of freedom on $SL(2,\mathbb{C})$ is defined, together with its Hamiltonian description. 
 
\end{abstract}

\newpage
\tableofcontents

\section{Introduction}
\label{intro}

Duality symmetries play a fundamental role in physics, relating different theories in many perspectives. One of the most fundamental in the context of String Theory is the so called T-duality \cite{giveon94, alvarez95, duff}, which is peculiar of  strings as extended objects and relates  theories defined on different target space backgrounds. The original notion of T-duality emerges in toric compactifications of the target background spacetime. The most basic example is provided by compactification of a spatial dimension on a circle of radius $R$. Here T-duality acts by exchanging momenta $p$ and winding numbers $w$, $p \leftrightarrow w$, while mapping  $R \rightarrow \frac{\alpha'}{R}$, with   $\alpha'$  the string fundamental length. This leads to a duality between string theories defined on different backgrounds but yielding the same physics, as it can be easily seen looking at the mass spectrum.

Interestingly, T-duality allows to construct new string backgrounds which could not be obtained otherwise, which are generally referred to as \textit{non-geometric backgrounds} (see for example \cite{plauschinn} for a recent review).\footnote{
By  non-geometric background it is  intended  a string configuration which cannot be described in terms of  Riemannian geometry. T-duality transformations are therefore needed  for gluing  coordinate patches, other than the usual diffeomorphisms and $B$-field gauge transformations.
}
Moreover,  it  plays an important role, together with S-duality and U-duality, in relating, through a web of dualities, the five superstring theories  which in turn appear as low-energy limits of a more general theory, that is,  M-theory. 

T-duality is certainly to be taken into account when looking at quantum field theory as low-energy limit of the  string action.  This has suggested since long \cite{duff, hull, tseytlin1, tseytlin2, siegel, lee} to look for a manifestly T-dual invariant formulation of the Polyakov  world-sheet action that has to be based on a doubling of the string coordinates in  target space. One relevant objective of this new action would be to obtain new indications for string gravity. This approach leads to Double Field Theory (DFT) 
with Generalised and Doubled Geometry furnishing the appropriate mathematical framework. In particular, DFT is expected to emerge as a low-energy limit of manifestly T-duality invariant string world-sheet. Then, Doubled Geometry is necessary to accommodate the coordinate doubling in target space. There is a vast literature concerning DFT, including its topological aspects and its description on group manifolds \cite{HZ,nunez, hullzw, blumen1, blumen2, HZ2, groot, pezzella3, copland1, park, berman2, berman1, hassler, pezzella2, pezzella1, bandos}. Recently, a   global formulation from higher Kaluza-Klein theory has been proposed in ref.  \cite{alfonsi}.

The kind of T-duality discussed so far belongs to a particular class, so called \textit{Abelian T-duality}, which is characterised by the fact that the generators of  target space duality transformations are Abelian,  while generating symmetries of the action only if they are Killing vectors of the metric \cite{buscher1, buscher2, rocekverlinde}. However, starting from Ref. \cite{ossa-quevedo}, it was realised that the whole  construction could be generalised to include the possibility that  one of the two isometry groups be non-Abelian.  This  is called \textit{non-Abelian}, or, more appropriately, \textit{semi-Abelian} duality. Although interesting, because it enlarges the possible geometries involved, the latter  construction is not really symmetric, as a duality would require. In fact, the dual model is typically missing some isometries which are required to go back to the original model by gauging. This means that one can map the original model to the dual one, but then it is not possible to go back anymore. This unsatisfactory feature is overcome with the introduction of \textit{Poisson-Lie T-duality} \cite{klimcik1,klimcik2,klimcik3}   (for some recent work to alternative approaches see \cite{bugdentesi, bugden19}). The latter represents a genuine generalisation,  since it does not require isometries at all, while  Abelian and non-Abelian cases can be obtained as particular instances. Recent results  on Poisson-Lie T-duality and its relation with para-Hermitian geometry and integrability, as well as low-energy descriptions, can be found in  \cite{klimcikdef, severa17, hassler17, jurco18, severa16, severa15, severa18, jurco19, marotta18, marotta19}. 

Symmetry under Poisson-Lie duality transformations is based on the concept of Poisson-Lie dual groups and Drinfel'd doubles. 
A Drinfel'd double is an even-dimensional Lie group $D$ whose Lie algebra $\mathfrak{d}$ can be decomposed into a pair of maximally isotropic subalgebras, $\mathfrak{g}$ and $\tilde{\mathfrak{g}}$, with respect to a non-degenerate ad-invariant bilinear form on $\mathfrak{d}$.  Lie algebras $ \mathfrak{g},\tilde{\mathfrak{g}}$ are dual as vector spaces, and endowed with compatible Lie structures. Any such triple, $(\mathfrak{d}, \mathfrak{g}, \tilde{\mathfrak{g}})$, is referred to as a Manin triple. If  $D, G,  \tilde{G}$ are the corresponding Lie groups, $G, \tilde G$ furnish an Iwasawa decomposition of $D$. The simplest example of Drinfel'd double is the cotangent bundle of any $d$-dimensional Lie group $G$, $T^* G \simeq G \ltimes \mathbb{R}^d$, which we shall call the \textit{classical double}, with trivial Lie bracket for the dual algebra $\tilde{\mathfrak{g}} \simeq \mathbb{R}^d$. In general, there may be many decompositions of $\mathfrak{d}$ into maximally isotropic subspaces (not necessarily subalgebras). The  set of all such decompositions plays the role of the modular space of field theories mutually connected by a T-duality  transformation. In particular, for the Abelian T-duality of the string on a $d$-torus, the Drinfel'd double is $D = U(1)^{2d}$ and its modular space is in one-to-one correspondence with $O(d,d;\mathbb{Z})$ \cite{klimcik3}. 

One can use Drinfel'd doubles to classify T-duality. Indeed, 
\begin{itemize}
\item Abelian doubles, characterised by Abelian algebras  $\mathfrak{g}, \tilde{\mathfrak{g}}$,   correspond to the standard Abelian T-duality;
\item semi-Abelian doubles, in which $\tilde{\mathfrak{g}}$ is Abelian, correspond to non-Abelian T-duality;
\item  non-Abelian doubles, which comprise all the other cases, correspond to the more general Poisson-Lie T-duality, where no isometries hold for either of the two dual models.
\end{itemize}

The appropriate geometric setting to investigate issues related to Poisson-Lie duality is that of dynamics on group manifolds. In this paper,  we consider in particular the $SU(2)$  Wess-Zumino-Witten (WZW) model in two space-time dimensions, which is a non-linear sigma model with the  group manifold of $SU(2)$ as target space, together with a topological cubic term. Needless to say, non-linear sigma models play an important role in many sectors of theoretical physics,  with applications ranging from the description of low energy hadronic excitations in four dimensions \cite{witten83, witten832}, to the construction of string backgrounds, like plane waves \cite{nappi, kehagias}, $AdS$ geometries \cite{MO,MO2,MO3, berkovits, gotz} or two-dimensional black hole geometries \cite{wittenbh}. Interesting examples of string backgrounds come from WZW models on non-semisimple Lie groups, that we will also consider in our work.  In the  context of   two-dimensional conformal field theories, gauged WZW models with coset target spaces are investigated since many years (see    \cite{witten91mk, chung} for early contributions).  Recently, non-linear sigma models  found new   applications in statistical mechanics, describing certain two-dimensional systems  at criticality \cite{robertsonpotts}, as well as in condensed matter physics, describing transitions for the integer quantum Hall effect \cite{zirnbauer}. 

From a theoretical point of view, non-linear sigma models represent  natural field theories on group manifolds, being intrinsically geometric. They play a   fundamental role  in  standard approaches to Poisson-Lie T-duality \cite{Alekseev1996, Sfetsos, Falceto, Bonechi} and  in the formulation of Double Field Theory  \cite{blumen1}. 
Once formulated on Drinfel'd doubles, such models allow for establishing  enlightening  connections with Generalized  Geometry (GG)  \cite{hitchin1, hitchin2, gualtieri:tesi}, by virtue of the fact that tangent and cotangent vector fields of the group manifold may be respectively related to the span of its Lie algebra and of the dual one. Locally, GG is based on replacing the tangent bundle $TM$ of a manifold $M$ with a kind of Whitney sum $TM \oplus T^*M$, a bundle with the same base space but fibres given by the direct sum of tangent and cotangent spaces, and the Lie brackets on the sections of $TM$ by the so called Courant brackets, involving vector fields and one-forms. Both the brackets and the inner products naturally defined on the generalised bundle are invariant under diffeomorphisms of $M$. More generally, a generalized tangent bundle is a vector bundle $E \to M$ enconded in the exact sequence $0 \to T^*M \to E \to TM \to 0$. This formal setting is certainly relevant  in the context of  DFT because it takes into account in a unified fashion vector fields, which generate diffeomorphisms for the background metric $G$ field, and one-forms, generating diffeomorphisms for the the background two-form $B$ field. In this framework Doubled Geometry plays a natural role in describing generalised dynamics on the tangent bundle $TD \simeq D \times \mathfrak{d}$, which encodes within a single action dually related models. 

In this paper we will follow an approach already proven to be successful for the Principal Chiral Model (PCM)\cite{marottapcm}, where  Poisson-Lie symmetries of the PCM with target space the manifold of $SU(2)$ are investigated. The guiding idea is  already  present in Ref.s \cite{MPV18, bascone, Vitale19} where  the simplest example of dynamics on a Lie group, the three-dimensional Isotropic Rigid Rotator, was considered as a  one-dimensional sigma model having $\mathbb{R}$ as source space and $SU(2)$ as target space. 
It is interesting to note that already in such a simple case, many aspects of Poisson-Lie T-duality can be exploited, and especially some relations with Doubled Geometry, although the model is too simple to exhibit manifest invariance.
In \cite{marottapcm}, the two-dimensional PCM on $SU(2)$ was considered, by means of  a one-parameter family of Hamiltonians and Poisson brackets,  all equivalent from the point of view of  dynamics.  
Poisson-Lie symmetry and a family of Poisson-Lie T-dual models were established.  Some connections with Born geometry were also made explicit. 

In this paper we will further extend the construction of the PCM by introducing a Wess-Zumino term, leading to a WZW model on $SU(2)$. We describe the model in the Hamiltonian approach with a pair of currents valued in the target phase space $T^*SU(2)$, which, topologically, is the manifold $S^3 \times \mathbb{R}^3$, while as a group it is the semi-direct product $SU(2) \ltimes \mathbb{R}^3$. An important feature is the fact that  as a symplectic manifold, $T^*SU(2)$ is symplectomorphic to $SL(2,\mathbb{C})$, besides being topologically equivalent.  Moreover,  both manifolds, $T^*SU(2)$ and $SL(2,\mathbb{C})$ are Drinfel'd doubles of the Lie group $SU(2)$ \cite{drinfeld1, drinfeld2, semenov, kossmann}, the former being the trivial one, what  we called classical double, which can be obtained from the latter via group contraction. 

The whole construction relies on a deformation of the affine current algebra of the model, the semidirect sum of the Kac-Moody algebra associated to $\mathfrak{su}(2)$ with  an Abelian algebra $\mathfrak{a}$, to the fully semisimple Kac-Moody algebra $\mathfrak{sl}(2,\mathbb{C})(\mathbb{R})$ \cite{R892, RSV93, R89}. The latter is a crucial step if one observes that the algebra $\mathfrak{sl}(2,\mathbb{C})$ has a bialgebra structure, with $\mathfrak{su}(2)$ and  $\mathfrak{sb}(2,\mathbb{C})$ dually related, maximal isotropic subalgebras.\footnote{ We denote with $\mathfrak{sb}(2,\mathbb{C})$ the  the Lie algebra of $SB(2,\mathbb{C})$, the Borel subgroup of $SL(2,\mathbb{C})$ of $2 \times 2$ complex valued upper triangular matrices with unit determinant and real diagonal.  By $\mathfrak{g}(\mathbb{R})$ we shall indicate the affine algebra  of maps $\mathbb{R} \to \mathfrak{g}$ that are sufficiently fast decreasing at infinity to be square integrable, what we will refer to as current algebra. } Current algebra deformation is also the essence of a Hamiltonian formulation of the classical world-sheet theory proposed in \cite{osten}. 

Starting from the one-parameter family of Hamiltonian models with algebra of currents homomorphic to $\mathfrak{sl}(2,\C)(\R)$, a further deformation is needed, in order to make the role of dual subalgebras completely symmetric. We show that such a deformation is possible, which does not alter the nature of the current algebra, nor the dynamics described by  the new Hamiltonian. In this respect, our findings will differ from existing results, such as $\eta$ or  $\lambda$ deformations of non-linear sigma models,   which represent true deformations of the dynamics  yielding to integrable models - recently, relations of these deformed models with Poisson-Lie T-duality have been found and worked out  in \cite{klimcikdef}.  We  end up with a  two-parameter family of models with the group $SL(2,\C)$ as target phase space. T-duality transformations are thus realised as $O(3,3)$ rotations in phase space. By performing an exchange of momenta with configuration space fields we obtain a new family of WZW models, with configuration space the group $SB(2,\C)$, which is dual to the previous one by construction.

The paper is organised as follows.
\noindent In Section \ref{sectwzwsu2} the Wess-Zumino-Witten model on the $SU(2)$ group manifold will be introduced with particular emphasis on   its Hamiltonian formulation and care will be payed to enlighten the Lie algebraic structure of   the Poisson brackets of fields.  The main purpose will be to illustrate the one-parameter deformation of the natural current algebra structure of the model to  
the affine Lie algebra associated to $\mathfrak{sl}(2,\mathbb{C})$ \cite{RSV93}. 

Section  \ref{sectplsym}  
is dedicated to Poisson-Lie symmetry in the Lagrangian and Hamiltonian context. While the former is standard and widely employed in the context of sigma models, we shall work out the Hamiltonian counterpart and verify its realisation within the model under analysis.
 In Section \ref{sectduality}
 a further parameter is introduced in the 
  current algebra 
  in such a way to make the role of the $\mathfrak{su}(2)$ and $\mathfrak{sb}(2,\mathbb{C})$ subalgebras symmetric, without modifying the dynamics. This is needed in order to have a manifest Poisson-Lie duality map, which reveals itself to be an $O(3,3)$ rotation in the target phase space $SL(2,\mathbb{C})$. Such a transformation leads  to a  two-parameter family of  models with  $SB(2,\mathbb{C})$ as target configuration space, which is dual to the starting family by construction. 
 
Independently from the previous Hamiltonian derivation,  in Section \ref{sectwzwsb2c} a WZW model on $SB(2,\mathbb{C})$ is  introduced  in the Lagrangian approach, together with  the corresponding string spacetime background. The model is interesting per se, because it is an instance of a  WZW  model with non-semisimple Lie group as target space, which exhibits classical conformal invariance. We overcome the intrinsic difficulties deriving from the absence of non-degenerate Cartan-Killing metric. However, the resulting dynamics does not seem to be related by a duality transformation to any of the models belonging to the parametric  family described above. We identify the problem as a topological obstruction and we show that in order to establish a connection  with any other of the models found,  a true deformation of the dynamics is needed, together with  a topological modification of the phase space. 

Finally, having understood what are the basic structures involved in the formulation of both the dually related  WZW families, in Section \ref{sectdoubledwzw} we introduce a generalised doubled WZW action  on the Drinfel'd double $SL(2,\mathbb{C})$ with doubled degrees of freedom. Its Hamiltonian description is presented and from it the Hamiltonian descriptions of the two submodels can be obtained by constraining the dynamics to coset spaces ${SU}(2)$ and ${SB}(2,\mathbb{C})$.
 
In Appendix \ref{sl2c} the mathematical setting of Poisson-Lie groups and Drinfel'd doubles is reviewed. In particular, the explicit construction of the Drinfel'd double group $SL(2,\mathbb{C})$ with respect to the Manin triple decomposition $(\mathfrak{sl}(2,\mathbb{C}), \mathfrak{su}(2), \mathfrak{sb}(2,\mathbb{C}))$ is presented with some detail, it being of central importance throughout the paper. 

Conclusions and Outlook are reported in the final Section \ref{sectconclusions}.

\section{The WZW model on $SU(2)$}
\label{sectwzwsu2}

The subject of this section is the Wess-Zumino-Witten model with target space the group manifold of $SU(2)$. First we review the model in the Lagrangian approach and then focus on its Hamiltonian formulation, the latter being more convenient for our purposes. 

 
 The main theme of the section is to describe the WZW model with an alternative canonical formulation in terms of a one-parameter current algebra deformation, based on  Ref.  \cite{RSV93}.  Such a richer structure  has several interesting consequences; some of them have already been investigated, such as quantisation \cite{RSV93} and integrability \cite{RSV96}, but in particular it paves the way to target space duality, presented in Section \ref{sectduality}. We follow the approach of \cite{marottapcm} where a similar analysis has been performed for the Principal Chiral Model.

\subsection{Lagrangian formulation}
\label{themodel}

Let $G$ be a semisimple connected Lie group and $\Sigma$ a $2$-dimensional oriented (pseudo) Riemannian manifold (we take it with Minkowski signature $(1,-1)$) parametrized by the coordinates $(t,\sigma)$.

The basic invariant objects we need in order to build a group-valued field theory are the left-invariant (or the right-invariant) Maurer-Cartan one-forms, which, if $G$ can be embedded in $GL(n)$, can be written explicitly as $g^{-1}dg \in \Omega^1(G) \otimes \mathfrak{g}$. 

Let us denote with $*$ the Hodge star operator on $\Sigma$, acting accordingly to the Minkowski signature as $*dt=d\sigma$, $*d\sigma=dt$. 

There is a natural scalar product structure on the Lie algebra of a semisimple Lie group, provided by the Cartan-Killing form and  denoted generically with the $\text{Tr}(\cdot, \cdot)$ symbol.

With this notation, we have the following
\begin{definition}
Let $\varphi: \Sigma \ni \left(t,\sigma\right) \rightarrow g \in G$ and denote $\varphi^*(g^{-1}dg)$ the pull-back of the Maurer-Cartan left-invariant one-form on $\Sigma$ via $\varphi$. The {Wess-Zumino-Witten model} is a non-linear sigma model described by the action
\begin{equation}
\label{wzwaction}
S=\frac{1}{4 \lambda^2} \int_{\Sigma} \text{Tr}\left[ \varphi^*\left(g^{-1} dg\right) \wedge * \varphi^*\left(g^{-1} dg \right)\right]+\kappa S_{WZ},
\end{equation}
with $S_{WZ}$  the  {Wess-Zumino term},
\begin{equation}
\label{wzdef}
S_{WZ}=\frac{1}{24 \pi}\int_{\mathcal{B}}\text{Tr}\left[ \tilde{\varphi^*}\left(\tilde{g}^{-1} d\tilde{g} \wedge \tilde{g}^{-1} d\tilde{g} \wedge \tilde{g}^{-1} d\tilde{g} \right)\right],
\end{equation}
where $\mathcal{B}$ is a 3-manifold whose boundary is the compactification of the original two-dimensional spacetime, while $\tilde{g}$ and $\tilde{\varphi}$ are  extensions of previous maps to the 3-manifold $\mathcal{B}$.
\end{definition}
It is always possible to have such an extension since one is dealing with maps $\varphi: S^2 \to G$. The latter  are classified by the second homotopy group $\Pi_2\left(G\right)$, which is well-known to be  trivial for  Lie groups. Thus, these maps are homotopically equivalent to the constant map, which can be obviously continued to the interior of the sphere $S^2$. Such an extension is not unique by the way, since  there may be many 3-manifolds with the same boundary. However, it is possible to show that the variation of the WZW action remains the same up to a constant term, which is irrelevant classically.  For the quantum theory, in order for  the partition function to  be single-valued, $\kappa$ is taken to be an integer  for compact Lie groups (this is the so called level of the theory), while for non-compact Lie groups there is no such a quantization condition.

For future convenience the action can be written explicitly as
\begin{equation}
S=\frac{1}{4 \lambda^2} \int_{\Sigma} d^2 \sigma \, \text{Tr}\left(g^{-1}\partial^{\mu}g g^{-1}\partial_{\mu}g \right)+\frac{\kappa}{24 \pi}\int_{\mathcal{B}} d^3 y \, \epsilon ^{\alpha \beta \gamma}\text{Tr}\left( \tilde{g}^{-1}\partial_{\alpha}\tilde{g}\tilde{g}^{-1}\partial_{\beta}\tilde{g}\tilde{g}^{-1}\partial_{\gamma}\tilde{g}\right).
\end{equation}
Note that although the WZ term is expressed as a three-dimensional integral, since $H \equiv \tilde g^{-1}d\tilde g ^{\wedge 3}$ is a closed 3-form, under the variation $ g \rightarrow  g+ \delta g$ (or more precisely $\varphi+\delta \varphi$) it produces  a boundary term, which is exactly an integral over $\Sigma$ since the variation of its Lagrangian density can be written as a total derivative. 
We have indeed
\begin{equation}
\delta S_{WZ}=\int_{\mathcal{B}}\mathcal{L}_{\tilde V_a}H=\int_{\mathcal{B}} d i_{\tilde V_a}H=\int_{\del  \mathcal{B}} i_{V_a}H,
\end{equation}
with $ {\del  \mathcal{B}} = \Sigma$, $\tilde V_a, V_a$ the infinitesimal generators of the variation over $ \mathcal{B}$ and $\Sigma$ respectively and $\mathcal{L}_{\tilde{V}_a}$ the Lie derivative along the vector field $V_a$. Then, its contribution to the equations of motion only involves the original fields  $\varphi$ on the source space $\Sigma$.

A remarkable property of the model is that its Euler-Lagrange equations may be rewritten as an equivalent system of  first order partial differential equations:
\begin{equation}
\label{eqmotexpl}
\partial_{t}A-\partial_{\sigma}J=-\frac{\kappa \lambda^2}{4 \pi}\left[A,J \right]
\end{equation}
\begin{equation}
\label{intcond}
\partial_t J-\partial_{\sigma}A=-\left[A,J \right]
\end{equation}
with 
\beqa 
A&=&\left(g^{-1}\partial_t g \right)^ie_i=A^i e_i, \\
J&=&\left(g^{-1}\partial_{\sigma} g \right)^ie_i=J^i e_i
\eeqa
Lie algebra valued fields (so called currents), $e_i \in \mathfrak{g}$,  
and 
  the usual physical boundary condition
\begin{equation}
\label{bcond}
\lim_{|\sigma| \rightarrow \infty} g(\sigma)=1,
\end{equation}
which makes the solution for $g$ unique. This boundary condition has also the purpose to one-point compactify the source space $\Sigma$.

At fixed $t$, the group elements satisfying this boundary condition form an infinite dimensional Lie group: $G(\mathbb{R}) \equiv \text{Map}(\mathbb{R},G)$, which is given by the smooth maps $g : \mathbb{R} \ni \sigma \to g(\sigma) \in G$ constant at infinity, with standard pointwise multiplication. 

The real line  may be  replaced by any smooth manifold $M$, of dimension $d$, so to have  fields in   $\text{Map}(M, G)$. The corresponding Lie algebra  $\mathfrak{g}(M) \equiv \text{Map}(M, \mathfrak{g})$ of maps $M \to \mathfrak{g}$ that are sufficiently fast decreasing at infinity to be square integrable (this is needed for the finiteness of the energy, as we will see) is the related  \textit{current algebra}. 

We will stick to the two-dimensional case from now on. 
Infinitesimal generators of the Lie algebra $\mathfrak{g}(\mathbb{R})$ can be obtained by considering the vector fields which generate the finite-dimensional Lie algebra $\mathfrak{g}$ and replacing ordinary derivatives with functional derivatives:
\begin{equation}
X_{i}(\sigma)={X_{i}}^{a}(\sigma) \frac{\delta}{\delta g^{a}(\sigma)},
\end{equation}
with Lie bracket
\begin{equation}
\label{liecurr}
\left[X_{i}(\sigma), X_{j}\left(\sigma^{\prime}\right)\right]={c_{i j}}^{k} X_{k}(\sigma) \delta^{d}\left(\sigma-\sigma^{\prime}\right),
\end{equation}
where $\sigma, \sigma' \in \mathbb{R}$. The latter  is $C^{\infty}(\R)$-linear and $\mathfrak{g}(\R) \simeq \mathfrak{g} \otimes C^{\infty}(\R)$. 

Let us now consider the target space $G=SU(2)$ and $\mathfrak{su}(2)$ generators $e_i=\sigma_i/2$, with  $\sigma_i$ the Pauli matrices, satisfying $[e_i, e_j]=i{\epsilon_{ij}}^k e_k$ and $\text{Tr}(e_i,e_j)=\frac{1}{2}\delta_{ij}$.

Eq.  (\ref{eqmotexpl}) can be easily obtained from the Euler-Lagrange equations for the action (\ref{wzwaction}).  Eq.  (\ref{intcond}) can be interpreted as an integrability condition for the existence of $g \in SU(2)$ such that $A=g^{-1}\partial_t g$ and $J=g^{-1}\partial_\sigma g$, and it follows from the Maurer-Cartan equation for the $\mathfrak{su}(2)$-valued one-forms $g^{-1}dg$.
This can be seen starting from the decomposition of the exterior derivative on the Maurer-Cartan left-invariant one-form:
\begin{equation*}
d\varphi^*\left(g^{-1}dg \right)=d\left(g^{-1}\partial_t g ~ dt+g^{-1}\partial_{\sigma}g~ d\sigma \right)=\left[- \partial_{\sigma}\left(g^{-1}\partial_t g\right) + \partial_t \left( g^{-1}\partial_{\sigma}g\right)\right]dt \wedge d\sigma,
\end{equation*}
and since
\begin{equation*}
\begin{aligned}
d\varphi^*\left(g^{-1}dg \right){} &=-\varphi^*(g^{-1}dg ) \wedge \varphi^*(g^{-1}dg) =-\left(g^{-1}\partial_t g g^{-1}\partial_{\sigma}g-g^{-1}\partial_{\sigma} g g^{-1}\partial_tg \right)dt \wedge d\sigma \\ & =-\left[g^{-1}\partial_t g, g^{-1}\partial_{\sigma}g\right]dt \wedge d\sigma,
\end{aligned}
\end{equation*}
Eq. (\ref{intcond}) follows.

To summarise,  the carrier space of Lagrangian dynamics can be regarded as the tangent bundle $TSU(2)(\mathbb{R}) \simeq (SU(2)  \ltimes \mathbb{R}^3)(\mathbb{R})$. It can be described  in terms of coordinates $(J^i,A^i)$, with $J^i$ and $A^i$ playing the role of left generalised configuration space coordinates and left generalised velocities respectively. In  the next section we will consider the Hamiltonian description, by replacing the generalised velocities $A^i$ with canonical momenta $I_i$ spanning the fibres of the cotangent bundle $T^* SU(2)(\mathbb{R})$.

For future convenience we close this section by introducing the form of the WZ term on $SU(2)$ in terms of the Maurer-Cartan one-form components:
\begin{equation}
S_{WZ}=\frac{1}{24 \pi}\int_{\mathcal{B}} d^3y \, \epsilon^{\alpha \beta \gamma} \tilde{A}_{\alpha }^i \tilde{A}_{\beta }^j \tilde{A}_{\gamma }^k {\epsilon_{ijk}} =\frac{1}{4\pi}\int_{\mathcal{B}} d^3y \, \epsilon^{\alpha \beta \gamma}\tilde{A}_{\alpha 1} \tilde{A}_{\beta 2} \tilde{A}_{\gamma 3},
\end{equation}
with $\tilde{A}_{\alpha}^i$ defined from $\tilde{\varphi}^*\left(\tilde{g}^{-1} d \tilde{g}\right)=\tilde{A}_{\alpha }^i dy^{\alpha}\, {e}_i  $.

\subsection{Hamiltonian description and deformed $\mathfrak{sl}(2,\mathbb{C})(\mathbb{R})$ current algebra} \label{hamdes}

The Hamiltonian description of the model is the one which mostly lends itself to the introduction of current algebras. 
The dynamics   is described by the following Hamiltonian
 \begin{equation}
\label{hamilt}
H=\frac{1}{4\lambda^2}\int_{\mathbb{R}} d\sigma \, \left(\delta^{ij}I_i I_j+\delta_{ij}J^i J^j \right)= \frac{1}{4\lambda^2}\int_{\mathbb{R}} d\sigma \, I_L ({\mathcal{H}^{-1}_0})_{LM} I_M
\end{equation}
and equal-time Poisson brackets
\begin{equation}
\label{Poissonalgwz}
\begin{aligned}
{} & \{I_i (\sigma), I_j(\sigma')\}=2\lambda^2\left[{\epsilon_{ij}}^k I_k(\sigma)+\frac{\kappa \lambda^2}{4 \pi} \epsilon_{ijk}J^k(\sigma)\right]\delta(\sigma-\sigma') \\ &
\{I_i (\sigma), J^j(\sigma')\}=2\lambda^2\left[{\epsilon_{ki}}^j J^k(\sigma)\delta(\sigma-\sigma')-\delta_i^j \delta'(\sigma-\sigma') \right] \\ &
\{J^i(\sigma),J^j(\sigma') \}=0
\end{aligned}
\end{equation}
which may be obtained from the action functional. For future reference  we have introduced in \eqn{hamilt} the double notation $I_L=(J^\ell, I_\ell)$ and the diagonal metric 
\begin{equation}
\label{diagmetric}
\mathcal{H}_{0}=\begin{pmatrix}
  \delta_{ij} & 0 \\
 0 & \delta^{ij}
 \end{pmatrix}.
\end{equation}
Momenta $I_i$ are obtained by Legendre transform from the Lagrangian.   Configuration space is the space of maps $SU(2)(\R)= \{g : \R\rightarrow SU(2)\}$,   with boundary condition  \eqn{bcond}, whereas the phase space $\Gamma_1$  is its cotangent bundle. As a manifold this is the product of $SU(2)(\R)$ with a vector space, its dual Lie algebra,  $ \mathfrak{su}(2)^*(\R)$, spanned by the currents $I_i$: 
\be
\Gamma_1=  SU(2)(\R) \times  \mathfrak{su}(2)^*(\R).
\ee
Hamilton equations of motion then read as:
\beqa
\partial_t I_j(\sigma)&=& \partial_{\sigma}J^k(\sigma) \delta_{kj} +\frac{\kappa \lambda^2}{4 \pi} {\epsilon_{jk}}^{\ell} I_{\ell}(\sigma) J^k(\sigma) \, , \label{eqwz}\\  
\partial_t J^j(\sigma)&=& \partial_{\sigma}I_k(\sigma) \delta^{kj} -{\epsilon^{j\ell}}_kI_{\ell}(\sigma)J^k(\sigma)  \, .
\label{intcondham}
\eeqa
Remarkably,  the Poisson algebra \eqn{Poissonalgwz} is homomorphic to $\mathfrak{c}_1$,  the semi-direct sum of the    Kac-Moody algebra associated to $SU(2)$ with the Abelian algebra $\R^3(\R)$:
\be
\mathfrak{c}_1=\mathfrak{su}(2)(\mathbb{R})\, \dot{\oplus} \, \mathfrak{a}.
\ee
Therefore,  the cotangent bundle  $\Gamma_1$ can be alternatively spanned by the conjugate variables $(J^j, I_j )$, with $J^j$ the left configuration space coordinates and $I_j$ the left momenta.

The energy-momentum tensor is traceless and conserved:
\begin{equation}\label{enmomtensor}
T_{00}=T_{11}=\frac{1}{4\lambda^2}\text{Tr}(I^2+J^2); \quad T_{01}=T_{10}=\frac{1}{2\lambda^2} \text{Tr}(IJ),
\end{equation}
so the model is conformally and Poincar\'e invariant, classically.

It has been shown in   Ref. \cite{RSV93}  that the current algebra $\mathfrak{c}_1$ may be deformed to a one-parameter family of fully non-Abelian algebras,   in such a way that the resulting brackets, together with a one-parameter family of deformed Hamiltonians, lead to an equivalent description of the dynamics. The new  Poisson  algebra was shown to be homomorphic to either $ \mathfrak{so}(4)(\R)$ or $ \mathfrak{sl}(2,\C)(\R)$, depending on the choice of  the deformation parameter. In  \cite{RSV93} the first possibility was investigated, while from now on  we shall choose the second option, for reasons that will be clear in a moment.  Accordingly, the cotangent space $\Gamma_1$ shall be replaced by a new one,  the set of $SL(2,\C)$ valued maps, $ \Gamma_2=  SL(2,\C) (\R)$. We refer to  \cite{RSV93} for  details about the deformation procedure, while 
hereafter we shall just state the result with a few steps which will serve our purposes. 
The new Poisson algebra will be indicated with  $\mathfrak{c}_2= \mathfrak{sl}(2,\C)(\R)$.  

\subsubsection{Deformation to the $\mathfrak{sl}(2,\mathbb{C})(\mathbb{R})$ current algebra}

Following the strategy already adopted in \cite{MPV18, marottapcm}, what is interesting for us is the occurrence of the group $SL(2,\C)$ as an alternative target phase space for the dynamics of the model. Indeed, $SL(2,\C)$ is the Drinfel'd double of  $SU(2)$, namely a group which can be  locally parametrised as a product of  $SU(2)$ with its properly defined dual, $SB(2,\C)$. The latter   is obtained by exponentiating the Lie algebra structure defined on the dual algebra of $\mathfrak{su}(2)$, under suitable compatibility conditions. Details of the construction are given in Appendix \ref{sl2c}. Since the role of the partner groups is symmetric,  we are going to see that  this shall allow to study Poisson-Lie duality in the appropriate  mathematical framework.

Before proceeding further, let us stress here that we are not going to deform the dynamics but only its target phase space description, and in particular its current algebra. This is completely different from the usual deformation approach followed for instance for integrable models. In that case one starts from a given integrable model, 
and then deforms it while trying to preserve the integrability property, but allowing for a modification of the physical content. In our case no deformation of the dynamics occurs. 

Inspired by Wigner-Inonu contraction of semisimple Lie groups, a convenient modification of the Poisson algebra $\mathfrak{c}_1$ which treats $I$ and $J$ on an equal footing is the following:
\begin{equation} \label{poal}
\begin{aligned}
{} & \{I_i (\sigma), I_j(\sigma')\}=\xi \left[{\epsilon_{ij}}^k I_k(\sigma)+a \, \epsilon_{ijk}J^k(\sigma)\right]\delta(\sigma-\sigma') \\ &
\{I_i (\sigma), J^j(\sigma')\}=\xi \left[\left({\epsilon_{ki}}^j J^k(\sigma) +b \, {\epsilon_i}^{jk} I_k(\sigma)\right) \delta(\sigma-\sigma')-\gamma\, \delta_i^j \delta'(\sigma-\sigma')  \right] \\ &
\{J^i(\sigma),J^j(\sigma') \}=\xi  \left[\tau^2 \epsilon^{ijk} I_k(\sigma)+ \mu \, {\epsilon^{ij}}_k J^k(\sigma)\right]\delta(\sigma-\sigma'),
\end{aligned}
\end{equation}
with $a,b,\mu, \xi,\gamma$ real parameters, while $\tau$ can be chosen either real or purely imaginary. Upon imposing that the equations of motion remain unchanged, it can be checked (see \cite{RSV93}) that it is sufficient to rescale the Hamiltonian by an overall factor, depending on $\tau$, according to 
\begin{equation}
\label{defhamilt}
H_{\tau}=\frac{1}{4\lambda^2(1-\tau^2)^2}\int_{\mathbb{R}} d\sigma \, \left(\delta^{ij}I_i I_j+\delta_{ij}J^i J^j \right)
\end{equation}
with the parameters obeying the constraints
\beqa
\xi&=& 2\lambda^2 \, (1-\tau^2) \label{xi}\\
a-b&=& \frac{\kappa \lambda^2}{4 \pi}(1-\tau^2)\\
\gamma&=&\-(1-\tau^2)
\eeqa
while $\mu$ is left arbitrary. 
In the limit $\tau, b, \mu \rightarrow 0$ we recover the standard description. 

For real $\tau$ the Poisson algebra \eqn{poal} is   isomorphic to $\mathfrak{so}(4)(\R)$ \cite{RSV93}, while for imaginary $\tau$ a more convenient choice of coordinates shall be done, which will make it evident the isomorphism with the $\mathfrak{sl}(2,\C)(\R)$ algebra. 
\\
Before doing that, let us shortly address the issue of space-time symmetries of the deformed model. The new formulation is still Poincar\'e and conformally invariant, although not being derived from the standard action principle. Indeed by following the same approach as in \cite{R89, RSV93} we obtain the new energy-momentum tensor, $\Theta_{\mu\nu}$,   by requiring that  
\be 
P= \int_{\mathbb{R}} d\sigma\, \Theta_{01}(\sigma)
\ee
and the Hamiltonian
\be
H= \int_{\mathbb{R}} d\sigma \, \Theta_{00}(\sigma)
\ee
generate space-time translations according to
\beqa
\frac{\del}{\del \sigma} \,I_k = \{P, I_k(\sigma)\}, \;\;\; && \frac{\del}{\del \sigma} J^k=\{P, J^k(\sigma)\} \nn\\
\frac{\del}{\del t} \,I_k = \{H, I_k(\sigma)\}, \;\;\; && \frac{\del}{\del t} J^k=\{H, J^k(\sigma)\}.
\eeqa
One finds 
\be
\Theta_{01}=\Theta_{10}=\frac{1}{4\lambda^2(1-\tau^2)^2}  \delta^{i}_j I_i J^j
\ee
\be
 \Theta_{00} = \frac{1}{4\lambda^2(1-\tau^2)^2} \left(\delta^{ij}I_i I_j+\delta_{ij}J^i J^j \right)  \,\, .
\ee
To obtain the remaining component of the energy-momentum  tensor,  we complete the  Poincar\'e algebra by introducing  the boost generator, $B$, which has to satisfy  the following Poisson brackets 
\be
\{H, B\}= P\;\;\; \{P, B\} = H.
\ee
The latter are verified by  
\be
B=-\frac{1}{2(1-\tau^2)}\int_{\mathbb{R}} d\sigma \, \sigma\, (\delta_{ij} J^i J^j + \delta^{ij} I_i I_j).
\ee
We thus compute the  boost transformations of $I$ and $J$, getting 
\be
\{I_\ell, B\}=2\lambda^2(1-\tau^2) \left( \sigma \frac{\del I_\ell}{\del t} + \delta_{\ell k} J^k\right), \;\;\; \{J^\ell, B\}= 2\lambda^2(1-\tau^2)\left(\sigma \frac{\del J^\ell}{\del t} + \delta^{\ell k} I_k\right)
\ee
namely, $I$ and $J$ transform as time and space components of a vector field.
Therefore the model is Poincar\'e invariant and the stress-energy tensor has to be conserved.  In particular 
\be
\frac{\del\Theta_{01}}{\del t}=\frac{\del\Theta_{11}}{\del \sigma}
\ee
 which yields $\Theta_{11}= \Theta_{00}$. 
\\
Conformal invariance is finally verified by computing  the algebra of the energy-momentum tensor, or, equivalently, by checking the  classical analogue of the Master Virasoro equation\footnote{The classical version of the Master Virasoro equation amounts to the following relations
\be \label{virasoro}
G^{AB}=  G^{AC} \Omega_{CD} G^{DB}, \;\;\; \tilde G^{AB}=  \tilde G^{AC} \Omega_{CD} \tilde G^{DB} \;\;\; 0=  \tilde G^{AC} \Omega_{CD} G^{DB}
\ee
with 
\be \Theta= \frac{1}{2}(\Theta_{00}+\Theta_{01})=G^{AB}M_A M_B\;\;\;   \tilde\Theta= \frac{1}{2}(\Theta_{00}-\Theta_{01})= \tilde G^{AB}M_A M_B \,\, 
\ee
$M_A=(J^a,I_a)$ and $\Omega_{AB}$ the matrix of central charges of the current algebra.
}(see for example \cite{HK89}). We do not repeat the calculation, performed in \cite{RSV93}, the only difference being the choice of $\tau$ as a real or imaginary parameter.
\\
\subsubsection{New coordinates}
It is convenient to introduce the real linear combinations
\begin{equation}
\label{transfsl}
\begin{aligned}
{} & S_i(\sigma)= \frac{1}{\xi(1-a^2 \tau^2)}\left[ I_i(\sigma)-a \delta_{ik} J^k(\sigma) \right] \,,  \\ &
B^i(\sigma)=\frac{1}{\xi (1-a^2 \tau^2)}\left[J^i(\sigma) -a \tau^2 \delta^{ik}I_k(\sigma)  \right] \, .
\end{aligned}
\end{equation}

On using the residual freedom for the parameters, we  choose $b=\mu=a \tau^2$ and $a=\frac{k\lambda^2}{4\pi}$, so that
\beqa
 \{S_i(\sigma), S_j(\sigma') \}&=&{\epsilon_{ij}}^kS_k(\sigma)\delta(\sigma-\sigma')+C \delta_{ij}\delta'(\sigma-\sigma') \label{sisi}\\ 
\{B^i(\sigma), B^j(\sigma') \} &=&\tau^2\epsilon^{ijk}S_k(\sigma)\delta(\sigma-\sigma')+\tau^2C\delta^{ij}\delta'(\sigma-\sigma') \label{bibi} \\
\{S_i(\sigma), B^j(\sigma') \} &=&{\epsilon_{ki}}^j B^k(\sigma)\delta(\sigma-\sigma')+C'{\delta_i}^j\delta'(\sigma-\sigma'),\label{sibi}
\eeqa
where we recognise rotations, $S_i$, and boosts, $B^i$, and
\be
C=
\frac{a}{\lambda^2(1-a^2\tau^2)^2}, \;\;\;\; 
C' =
-\frac{(1+a^2\tau^2)}{2 \lambda^2 (1-a^2\tau^2)^2}
\ee
the central charges.
Note that both transformations and Poisson algebra are consistent and non-singular in the limit $\tau \rightarrow 0$.
As an intermediate step, it is convenient to write the Hamiltonian in terms of $S$ and $B$. By replacing
\begin{equation}
\label{changebas1}
\begin{aligned}
{} & I_i(\sigma)= \xi\left[ S_i(\sigma)+a \delta_{ik} B^k(\sigma) \right] \\ &
J^i(\sigma)=\xi \left[B^i + a \tau^2 \delta^{ik} S_k(\sigma) \right]
\end{aligned}
\end{equation}
in Eq. (\ref{defhamilt}), it is easy to obtain:
\begin{equation}
\label{sbhamtau}
H_{\tau}=\lambda^2 \int_{\mathbb{R}} d\sigma \Big[ \left(1+a^2 \tau^4 \right)\delta^{ij}S_i S_j+\left(1+a^2 \right)\delta_{ij}B^i B^j+2 a\left(1+\tau^2 \right){\delta^i}_j S_i B^j \Big],
\end{equation}
where we suppressed  the $\sigma$-dependence of fields for the sake of notation.
The equations of motion in terms of the new generators $S$ and $B$ read as:
\begin{equation}
\label{eomsb}
\begin{aligned}
{} & \partial_t S_k=- \frac{a(1-\tau^2)}{1-a^2\tau^2} \partial_{\sigma} S_k+ \frac{1-a^2}{1-a^2 \tau^2} \delta_{pk} \partial_{\sigma}B^p \, ,
 \\ &
\partial_t B^k=\frac{1- a^2 \tau^4 }{1-a^2 \tau^2} \delta^{kp}\partial_{\sigma}S_p+ \frac{a(1-\tau^2)}{1-a^2 \tau^2} \partial_{\sigma}B^k- \xi (1-a^2\tau^2)  {\epsilon^{kp}}_q S_p B^q \, .
\end{aligned}
\end{equation}
Note that now non-diagonal terms appear in the Hamiltonian, which are zero not only for $a =0$ (i.e. without WZ term), which corresponds to the Principal Chiral Model \cite{marottapcm}, but also for $\tau=\pm i$, $a\ne 0$.

Our next goal  is to make explicit the bialgebra structure of the Poisson algebra \eqn{sisi}-\eqn{sibi}, according to the decomposition   $\mathfrak{sl}(2,\mathbb{C})= \mathfrak{su}(2) \Join \mathfrak{sb}(2,\mathbb{C})$ (see Appendix \ref{sl2c}). To this, another linear transformation  of the generators is needed.
We leave the $S_i$  unchanged since, according to Eq. \eqn{sisi}, they already span the $\mathfrak{su}(2)$ algebra
  and transform the $B^i$ generators as follows:
\begin{equation}
K^i(\sigma)=B^i(\sigma)\, +\,  i\tau \epsilon^{i \ell 3}S_{\ell}(\sigma) \, .
\end{equation}
The new generators span the affine algebra associated with the Lie algebra $\mathfrak{sb}(2,\C)$, as can be easily checked by computing their Poisson brackets, which read as:
\begin{equation}
\{K^i(\sigma), K^j(\sigma') \}=i\tau{f^{ij}}_k K^k(\sigma)\delta(\sigma-\sigma')+C\tau^2 h^{ij} \delta'(\sigma-\sigma'), \label{kiki}
\end{equation}
with ${f^{ij}}_k=\epsilon^{ij \ell} \epsilon_{\ell 3 k}$ the  structure constants of $\mathfrak{sb}(2,\mathbb{C})$ and 
\be
h^{ij}=\delta^{ij}+{\epsilon}^{ip3}\delta_{pq} \epsilon^{jq3}
\ee
a non-degenerate metric in  $\mathfrak{sb}(2,\mathbb{C})$ defined  in Eq.  (\ref{sbmetric}).
With similar calculations for the mixed bracket we find:
\begin{equation}
\begin{aligned}
\{S_i(\sigma) ,K^j(\sigma') \;\}=\left[{\epsilon_{ki}}^j K^k(\sigma)  +i\tau {f^{jk}}_i  S_k(\sigma)  \right] \delta(\sigma-\sigma')+\left( C'{\delta_i}^j- i\tau C{\epsilon_i}^{j3} \right)\delta'(\sigma-\sigma').\label{siki}
\end{aligned}
\end{equation}

The Poisson algebra described by Eqs. \eqn{sisi}, \eqn{kiki}, \eqn{siki}  is a bialgebra,  isomorphic to $\mathfrak{c}_2$, with its maximal subalgebras clearly identified as $\mathfrak{su}(2)(\R)$ and $\mathfrak{sb}(2,\C)(\R)$.

By substituting  
\begin{equation}
\label{changebas2}
B^i(\sigma)=K^i(\sigma)- i\tau \epsilon^{i \ell 3} S_{\ell}(\sigma)
\end{equation}
the Hamiltonian is rewritten in terms of the new generators as
\begin{equation}
\begin{aligned}
{} & H_{\tau}= \lambda^2 \int_{\mathbb{R}} d\sigma \Big\{S_i S_j \left[(1+a^2 \tau^4)\delta^{ij}-\tau^2(1+a^2){\epsilon}^{ip3} \delta_{pq}\epsilon^{jq3} \right] \\ & +K^i K^j (1+a^2)\delta_{ij}+S_i K^j\left[ 2a(1+\tau^2)\delta^{ip}+ 2i\tau(1+a^2){\epsilon}^{ip3}\right]\delta_{pj}  \Big\}.
\end{aligned}
\end{equation}
Let us notice that the model remains conformally invariant, because we have only performed linear transformations of the current algebra generators.

In compact form the Hamiltonian reads, 
\be
H_{\tau}=\lambda^2 \int_{\mathbb{R}} d\sigma \, S_I \left(\mathcal{M}_{\tau} \right)^{IJ} S_J,
\ee
where we have introduced  the  doubled notation $S_I \equiv (K^i, S_i)$, and the  generalised metric $\mathcal{M}_{\tau}(a)$, given by
\begin{equation}
\label{taudefham}
\mathcal{M}_{\tau}=\begin{pmatrix}
  (1+a^2 \tau^4)\delta^{ij}-\tau^2(1+a^2){\epsilon}^{ip3}\delta_{pq} \epsilon^{jq3} &  \left[i\tau(1+a^2){\epsilon}^{ip3} +a(1+\tau^2)\delta^{ip}\right]  \delta_{pj}\\
  \delta_{ip}\left[-i\tau(1+a^2)\epsilon^{pj3}+a(1+\tau^2)\delta^{pj} \right]& (1+a^2)\delta_{ij}
 \end{pmatrix}.
\end{equation}
Let us analyse the latter in more detail, as a function of the parameters $a,\tau$. For $a=0$ we retrieve the one-parameter family associated to the  PCM,  studied  in \cite{marottapcm}, and $\mathcal{M}_{\tau}(a=0)$  can be checked to be   an  $O(3,3)$ matrix, namely, $\mathcal{M}_{\tau} \eta \mathcal{M}_{\tau}=\eta$. For $a\ne 0$ the metric is not  $O(3,3)$ in general, but it could be for specific  values of the parameters; for example, it becomes proportional to an $O(3,3)$ matrix for $\tau=\pm i$. In particular for $\tau=-i$  we find  $\mathcal{M}_{-i}(a) = \left(1+a^2 \right)\mathcal{H}^{-1}$,  where 
\begin{equation}
\mathcal{H}^{-1}=\begin{pmatrix}
  h^{ij} & + {\epsilon}^{ip3}\delta_{pj}\\
  -\delta_{ip}{\epsilon}^{pj3} & \delta_{ij}
 \end{pmatrix}
\end{equation}
 is the inverse of the Riemannian metric (\ref{riemannianmetric}) which has been defined in Appendix \ref{sl2c} in the Lie algebra $\mathfrak{sl}(2,\mathbb{C})$.   

Let us summarise the main results of this section. The WZW model with target configuration space $SU(2)$ has been described in terms of a  one-parameter family of Hamiltonians $H_\tau$
\begin{equation}\label{hamsk}
H_{\tau}=\lambda^2 \int_{\mathbb{R}} d\sigma \, S_I \left(\mathcal{M}_{\tau} \right)^{IJ} S_J
\ee
with Poisson brackets realising the  non-compact current algebra  $\mathfrak{c}_2$
\beqa
\label{poissontaudefdrinf}
&& \{S_i(\sigma),\, S_j(\sigma') \; \}={\epsilon_{ij}}^kS_k(\sigma)\delta(\sigma-\sigma')+C \delta_{ij}\delta'(\sigma-\sigma') \nonumber\\
&& \{K^i(\sigma), K^j(\sigma') \}= i\tau{f^{ij}}_k K^k(\sigma)\delta(\sigma-\sigma')+C\tau^2 h^{ij} \delta'(\sigma-\sigma') \\
&&\{S_i(\sigma) ,K^j(\sigma') \;\}=\left[{\epsilon_{ki}}^j K^k(\sigma)  +i\tau {f^{jk}}_i  S_k(\sigma)  \right] \delta(\sigma-\sigma')+\left( C'{\delta_i}^j- i\tau C{\epsilon_i}^{j3} \right)\delta'(\sigma-\sigma')\nonumber
\eeqa
and 
\beqa\label{KItran}
K^i&=& \frac{1}{\xi(1-a^2\tau^2)}\left( (\delta_k^i-i\tau a \epsilon^{i\ell 3 }\delta_{\ell k})J^k + (\epsilon^{i k3}-a \tau^2 \delta^{ik})I_k\right) \nn\\
S_i &=&= \frac{1}{\xi(1-a^2\tau^2)}\left(I_i-a \delta_{ik} J^k\right).
\eeqa
The Poisson algebra, thanks to the choice performed for the generators,  reveals  a bialgebra structure (see Appendix \ref{sl2c}),  with central terms. 
It is interesting to note that the central terms of the brackets for the $\mathfrak{su}(2)$ and $\mathfrak{sb}(2,\mathbb{C})$ generators entail the metrics of the respective algebras, obtained directly from the generalised metric on $\mathfrak{sl}(2,\mathbb{C})$ in (\ref{riemannianmetric}).

The alternative canonical formulation which has been presented  here has some interesting features in relation with quantisation \cite{RSV93} and integrability \cite{RSV96}.  In the following we will exploit the bialgebra formulation to analyse the symmetries of the model under Poisson-Lie duality.

\section{Poisson-Lie symmetry}
\label{sectplsym}
In a field theory context Poisson-Lie symmetry \cite{klimcik1,klimcik2,klimcik3, bugdentesi} is usually introduced as a deformation of standard isometries of two-dimensional non-linear sigma models on pseudo-Riemannian manifolds, in the Lagrangian approach. To make contact with the existing literature let us therefore summarise the main aspects. We use here local, light-cone coordinates, to adhere to common approach. 
\begin{definition}
Let $X^i: \Sigma \to {M}$, where $(\Sigma, h)$ is a $2$-dimensional oriented pseudo-Riemannian manifold,  the so called \textit{source space} (or worldsheet) with metric $h$ and $({M}, g)$ a smooth manifold, the so  called \textit{target space} (or background), equipped with a metric $g$ and a $2$-form $B$. Let ${M}$ admit at least a free action of a Lie group $G$ \footnote{In general the action is only required to be free. If it is also transitive, the model takes the name of Principal Chiral Model and the target space is diffeomorphic with the group itself, as it is the case in this paper.}. A 2-dimensional \textit{non-linear sigma model} can be defined by the following action functional:
\begin{equation}
S=\int_{\Sigma} d z d \bar{z} \, E_{i j} \partial X^{i} \bar{\partial} X^{j},
\end{equation}
with the generalised metric $E_{ij}=g_{ij}+B_{ij}$.
\end{definition}
Suppose that the group $G$ acts freely from the right, then the infinitesimal generators of the right action are the left-invariant vector fields $\{{V_a}\}$,
satisfying 
\begin{equation}
[V_a, V_b]={f_{ab}}^c V_c
\end{equation}
with ${f_{ab}}^c$ the structure constants of $\mathfrak{g}$.
Under an  infinitesimal variation of the fields
\begin{equation}
\delta X^{i}=V_{a}^{i} \epsilon^{a},
\end{equation}
with $\epsilon$ the infinitesimal parameters of the transformation, the variation of the action reads as:
\begin{equation*}
\delta S=\int_{\Sigma} d z d \bar{z} \, \mathcal{L}_{V_{a}}\left(E_{i j}\right) \partial X^{i} \bar{\partial} X^{j} \epsilon^{a}-\int_{\Sigma} d z d \bar{z}\left[\partial\left(V_{a}^{i} E_{i j} \bar{\partial} X^{j}\right)+\bar{\partial}\left(V_{a}^{i} E_{j i} \partial X^{j}\right)\right] \epsilon^{a},
\end{equation*}
where  $\mathcal{L}_V$ denotes the Lie derivative along the vector field $V$.
Using the fact that 
\begin{equation*}
d z d \bar{z}\left[\partial\left(V_{a}^{i} E_{i j} \bar{\partial} X^{j}\right)+\bar{\partial}\left(V_{a}^{i} E_{j i} \partial X^{j}\right)\right]=d\left(V_{a}^{i} E_{i j} \bar{\partial} X^{j} d \bar{z}-V_{a}^{i} E_{j i} \partial X^{j} d z\right),
\end{equation*}
we are left with
\begin{equation} \label{dela}
\delta S=\int_{\Sigma} d z d \bar{z} \, \mathcal{L}_{V_{a}}\left(E_{i j}\right) \partial X^{i} \bar{\partial} X^{j} \epsilon^{a}-\int_{\Sigma} d J_{a} \epsilon^{a},
\end{equation}
with 
\be
\label{ja}
J_{a}={V_{a}}^{i}\left(E_{i j} \bar{\partial} X^{j} d \bar{z}-E_{j i} \partial X^{j} d z\right)
\ee
 the Noether 
one-forms associated to the group transformation.
If the action functional  is  required  to be invariant, $\delta S$ has to be zero. Usually a stronger requirement is applied, namely that the target-space geometry be invariant as well. This entails separately the invariance of the metric and of the $B$-field,  that is $\mathcal{L}_{V_{a}}\left(E_{i j}\right) =0$, $\mathcal{L}_{V_{a}}\left(B_{i j}\right) =0$.  The  two-form $B$ could   be put to zero to start with. Hence, under these assumptions, the  symmetry group is a group of isometries and the generators are Killing vector fields. If this is the case, from \eqn{dela} we derive that the Noether one-forms are closed
\be
d J_{a} =0
\ee
hence, they are locally exact,
\be
J_{a}= d\widetilde X_a.
\ee 
 In particular, if the symmetry group is  Abelian, one can always find a frame where ${V_a}^i={\delta_a}^i$. In this case Abelian T-duality can be obtained by exchanging $X^i$ with the dual coordinates $\widetilde{X}_i$  and the Bianchi identity $d^2 \tilde{X}_i=0$ then leads to the equations of motion for the original theory. 
 Notice that, because of the definition of the one-forms $J_a$, the functions $\widetilde X_a$ take value in the tangent space at ${M}$, namely, they are velocity coordinates. Therefore, the symmetry of the model under target-space duality transformation amounts to the exchange of target space coordinates $X^i$ with velocities $\tilde X_i$,   the generators of the symmetry are Killing vector fields and T-duality is along directions of isometry. 
 
 In the case in which the symmetry group of the action is non-Abelian, but still an isometry, one refers to non-Abelian (or, better, semi-Abelian)  T-duality, with Noether currents satisfying Abelian Maurer-Cartan equations.  


However, this whole construction can be generalised: suppose the Noether current one-forms are not closed but satisfy instead a Maurer-Cartan equation 
\begin{equation}
\label{noethermaurercartan}
d J_{a}=\frac{1}{2} \tilde{f}_{a}^{\, \, \,b c} J_{b} \wedge J_{c}
\end{equation}
being $\tilde{f}_{a}^{\, \, \,b c}$ the structure constants of some Lie algebra $\tilde{\mathfrak{g}}$ not yet specified. 
Using Eq.  (\ref{noethermaurercartan}), by imposing invariance of the action, Eq.  \eqn{dela} yields 
\begin{equation*}
\int_{\Sigma} d z d \bar{z} \, \mathcal{L}_{V_{a}}\left(E_{i j}\right) \partial X^{i} \bar{\partial} X^{j} \epsilon^{a}=\int_{\Sigma} \frac{1}{2} \tilde{f}_{a}^{\, \, \,b c} J_{b} \wedge J_{c} \, \epsilon^{a}.
\end{equation*}
From Eq. \eqn{ja}, it is straightforward to obtain
\begin{equation*}
J_{b} \wedge J_{c}=-2 V_{b}^{m} V_{c}^{l} E_{n m} E_{l k} \partial X^{n} \bar{\partial} X^{k} d z d \bar{z},
\end{equation*}
and finally 
\begin{equation}
\mathcal{L}_{V_{a}} E_{i j}=-\tilde{f}_{a}^{ \, \, \,b c} V_{b}^{k} V_{c}^{l} E_{i k} E_{l j}.
\end{equation}
The latter relation reveals that in order for the action to be invariant, the generators not only can be non-Abelian, but they do not have to be isometries (one can still find the standard isometry case if the algebra of Noether currents,  $\tilde {\mathfrak{g}}$,  is Abelian, so that the Lie derivative is again vanishing).  

If this is the case, we say that the sigma model is \textit{Poisson-Lie symmetric} (see def. \eqn{defA1} of Poisson-Lie group). Indeed, 
from the Lie algebra  condition
\begin{equation}
\left[\mathcal{L}_{V_{a}}, \mathcal{L}_{V_{b}}\right] E_{i j}={f_{a b}}^{c} \mathcal{L}_{V_{c}} E_{i j}
\end{equation}
the following compatibility condition for the pair structure constants follows:
\begin{equation}\label{compaco}
{\tilde{f}^{\, \, \,m c}}_a {f_{d m}}^b-{\tilde{f}^{\, \, \,m b}}_a {f_{d m}}^c-{\tilde{f}^{\, \, \,m c}}_d {f_{a m}}^b+{\tilde{f}^{\, \, \,m b}}_d {f_{a m}}^c-{\tilde{f}^{\, \, \,b c}}_m {f_{d a}}^m=0,
\end{equation}
which is exactly the compatibility condition in (\ref{bialgebracond}), in order for two algebras $\mathfrak{g}, \tilde{\mathfrak{g}}$ (which are dual as vector spaces), concur to define a bialgebra, $\mathfrak{d}$, whose underlying vector space is the direct sum of the former. Equivalently, Eq. \eqn{compaco} is nothing but the compatibility condition \eqn{compcondpl} between  Poisson and group structure of a Poisson-Lie group. 

The triple ($\mathfrak{d}, \mathfrak{g},\tilde{\mathfrak{g}})$  is   associated to the starting sigma model but,  since the construction of the bialgebra structure is completely symmetric, one can expect to formulate a model associated to the same triple  by swapping role of the subalgebras $\mathfrak{g}, \tilde{\mathfrak{g}}$ and that will be the \textit{Poisson-Lie dual sigma model}, defined by
\begin{equation}
\mathcal{L}_{\tilde{V}_{a}} \tilde{E}_{i j}=-{f_{a}}^{b c} \tilde{V}_{b}^{k} \tilde{V}_{c}^{\ell} \tilde{E}_{i k} \tilde{E}_{\ell j}
\end{equation}
where all fields with $\tilde\,$ refer to the dual model. 

On introducing the group $D$ which corresponds to the exponentiation of the bialgebra $\mathfrak{d}$, 
equivalently we can say that a sigma model is of Poisson-Lie type if the target space is a coset space $D/G$, where 
$G$ indicates one of its component groups in a chosen polarisation. Its dual will be defined on the target coset $D/\tilde{G}$. The group $D$ is the Drinfel'd double and ($\mathfrak{d}, \mathfrak{g},\tilde{\mathfrak{g}})$ is a Manin triple. $G$, $\tilde G$ are dual groups. Mathematical details  can be found in Appendix \ref{sl2c}.

Since Poisson-Lie T-duality is a generalisation of Abelian and semi-Abelian T-dualities,  T-dualities may be classified in terms of the types of Manin triple underlying the sigma model structure:
\begin{itemize}
\item Abelian doubles correspond to standard Abelian T-duality. The Drinfel'd double is Abelian, with Lie algebra $\mathfrak{d}=\mathfrak{g} \oplus \tilde{\mathfrak{g}}$ with the  algebra $\mathfrak{g}$  and its dual both Abelian;
\item Semi-Abelian doubles (i.e. $\mathfrak{d}= \mathfrak{g} \, \dot{\oplus} \, \tilde{\mathfrak{g}}$, with $\mathfrak{g}$ non-Abelian, $\tilde{\mathfrak{g}}$ Abelian and $\dot{\oplus}$ a semi-direct sum) correspond to non-Abelian T-duality between an isometric and a non-isometric sigma model; 
\item Non-Abelian doubles (all the other possible cases) correspond to Poisson-Lie T-duality. Here no isometries hold for either of the two dual models.
\end{itemize}

The notion of Poisson-Lie symmetry can also be formulated in the Hamiltonian formalism \cite{klimcik96, marottapcm, bascone, stern98, stern99}.   We may state  the following
\begin{definition}
Let $({M},\omega)$ be a symplectic manifold admitting a right action ${M} \times G \to {M}$ of $G$ on ${M}$, and let   $V_a \in \mathfrak{X}({M}), a=1,\dots d$ be the vectors fields which  generate the action, with $d={\rm dim}\, \mathcal{\mathfrak{g}}$.
If $\mathcal{L}_{V_a } \omega\ne 0$ but 
\begin{equation}
\label{plsymham}
 i_{V_a} \omega= \tilde {\theta}^a,
\end{equation}
with $\tilde {\theta^a}$  left(right)-invariant one-forms of the dual group $\tilde G$ and $i_V$ the interior derivative  along $V$,  we say that a dynamical system with phase space $({M}, \omega)$ is Poisson-Lie symmetric with respect to $G$ if its Hamiltonian is invariant\footnote{However, it should be sufficient  to require that the vector fields $V$ generate symmetries of the equations of motion, not necessarily of the Hamiltonian, which is, as the symplectic form, an auxiliary structure.} 
\begin{equation}
\mathcal{L}_V H=0.
\end{equation}
\end{definition}
For future convenience,  Eq. \eqn{plsymham} can be equivalently formulated according to
\be \label{MCeq}
\mathcal{L}_{V_a } \omega=- \frac{1}{2} {f^a}_{bc}\tilde {\theta}^b\wedge \tilde {\theta}^c
\ee
which, contracted with dual vector fields, yields $\mathcal{L}_{V_a } \omega (\tilde X_b, \tilde X_c)= - {f^a}_{bc}$.
  Finally, let us notice that Poisson-Lie symmetry may be stated in terms of the Poisson bi-vector field $\Pi$   by saying that a dynamical system with target space  a Poisson manifold possesses Poisson-Lie symmetry under the action of a Lie group $G$ if the Hamiltonian (or its equations of motion) is invariant and the bivector field $\Pi$ together with the infinitesimal generators of the symmetry implicitly defines one-forms of the dual group according to
\begin{equation} \label{poili}
V_a= \Pi (\tilde{\theta}^a).
\end{equation}

\subsection{Poisson-Lie symmetry of the WZW model}
Before looking explicitly at the dual models, we address the Poisson-Lie symmetry  of the one-parameter family of WZW models described by the Hamiltonian \eqn{hamsk}, and Poisson brackets \eqn{poissontaudefdrinf},  adapting  the definition given above to our setting. 

The analysis follows very closely the one performed  in \cite{marottapcm} for the Principal Chiral Model. 

Keeping the interpretation of $(K^i,S_i)$ as target phase space 
coordinates with $K^i$ and $S_i$ respectively base and fibre coordinates, one can associate Hamiltonian vector fields to $K^i$ by means of 
\begin{equation}
X_{K^i} \coloneqq  \{\cdot, K^i \}.
\end{equation}
The fields so defined obey by construction  the Lie algebra relations
\begin{equation}
\left[X_{K^{i}}, X_{K^{j}}\right]=X_{\left\{K^{i}, K^{j}\right\}}=i \tau {f^{ij}}_k X_{K^{k}},
\end{equation}
inherited from the  non-trivial Poisson structure (\ref{poissontaudefdrinf}). They span the Lie algebra $\mathfrak{sb}(2,\mathbb{C})$ and, in the  limit $\tau \to 0$, reproduce the original Abelian structure of the $\mathfrak{su}(2)$ dual. Because of their definition they satisfy
\be
\omega (X_{K^{j}},X_{K^{k}} )= \{K^j, K^k\}= {f^{jk}}_i K^i.
\ee 
Moreover, we may define dual one-forms in the standard way, $\alpha_j : \alpha_j(X_{K^{k}}) = \delta_j^k$, which   satisfy the Maurer-Cartan equation 
\be\label{MCe}
d\alpha_i(X_{K^{j}}, X_{K^{k}})=-\alpha_{i}([X_{K^{j}}, X_{K^{k}}])=-{f^{j k}}_i.
\ee
The latter, being basis one-forms of the dual algebra,  can be identified with  basis generators of $\mathfrak{su}(2)$, $\alpha_i\rightarrow V_i$.
By inverting Eq. \eqn{MCe}  one can check {that} this is indeed the Poisson-Lie condition  stated in Eq. \eqn{MCeq}.

 \subsubsection{$B$ and $\beta$ T-duality transformations}
It was already noticed in \cite{marottapcm} that the one-parameter family of Principal Chiral Models obtained from deformation of the target phase space could be recognised as a family of Born geometries, generated by $\beta$ T-duality transformations. The situation for the WZW model is  more involved. 
\\
Starting from the generalised vector $(J^i, I_i)$, which obeys the Poisson algebra \eqn{Poissonalgwz}, we have performed a series of transformations, ending up with a new generalised vector, $(K^i, S_i)$, satisfying the Poisson algebra \eqn{poissontaudefdrinf}, while describing the same dynamics. These transformations are therefore symmetries, which can be partially recast in the form of  $\beta$-transformations as follows. 
\\
According to \cite{lust-osten18, Deser15}, given a generalised vector field on the target space,    $(X, \omega)$, a 
 $\beta$-transformation in the context of Poisson-Lie groups is a T-duality transformation of the algebra of currents $\phi: \mathfrak{d}(\R)\rightarrow \mathfrak{d}(\R)$,  which may be represented as 
\be
(X, \omega)\rightarrow (X+ i_\omega \beta, \omega ) \;\;\; \beta\in \Gamma (\Lambda^2 TM)
\ee
with $\beta$ a bivector field. As dual  to the latter, another $T$-duality transformation is a $B$-transformation, given by 
\be
(X, \omega)\rightarrow (X, \omega + i_X B) \;\;\; B\in \Gamma (\Lambda^2 T^*M)
\ee
with $B$ a two-form.
Besides, there are other T-duality transformations, such as  factorised transformations, which may be rephrased  in the same setting as linear transformations in the bialgebra of generalised vector fields of the target space.
\\
In this perspective  let us see how 
the transformation \eqn{KItran} can be reformulated. 
Differently from the PCM model without WZW term, we need to split the transformation in two steps.  We first perform 
\be
(J^i, I_i)\rightarrow (\tilde J^i, \tilde I_i)= (J^i - a \tau^2\delta^{ik} I_k,\, I_i- a \delta_{ik} J^k)
\ee
which is a generalised linear transform of the kind
\be
(X, \omega) \rightarrow (X+C(\omega), \omega+ \tilde C (X))
\ee
with $C=-a  \tau^2\delta^{ij}\xi_i\otimes \xi_j$, $\tilde C= - a \delta_{ij} \xi^{*i} \otimes \xi^{*j} $, \,  $\{ \xi_i \}, \{ \xi^{*i} \},$  basis of vector fields and dual one-forms on $M$, and $\tilde C= - C^{-1}$   for $i\tau= 1/a$. 
We thus perform a $\beta$-transformation:
\be
(\tilde J^i, \tilde I_i)\rightarrow (\tilde J^i +i\tau \epsilon^{i\ell 3} \tilde I_\ell, \tilde I_i)\equiv (K^i, S_i).
\ee
The generalised metric \eqn{taudefham} may be obtained from the diagonal metric $\mathcal{H}_0$ applying the above transformations accordingly. 
Therefore, the one-parameter family of WZW models introduced in previous sections can be regarded as  a sequence of $B$- and $\beta$-transformations. For $a=0$, we resort to the PCM considered in \cite{marottapcm}, where the one-parameter family is a family of Born geometries related by pure $\beta$-transformations.

\section{Poisson-Lie T-duality}
\label{sectduality}
In order to investigate  duality transformations within the current algebra  which has been obtained at the end of Sec. \ref{hamdes}, we need to make the 
role of   the two subalgebras $\mathfrak{su}(2)(\R)$ and $\mathfrak{sb}(2,\mathbb{C})(\R)$ completely symmetric. To this, we shall introduce a further parameter in the current algebra, so to get  a {\it two-parameter} formulation of the  WZW model. As a result,  T-duality transformations will be realised  as  simple $O(3,3)$ rotations  in the  target phase space $SL(2,\mathbb{C})(\mathbb{R})$ and the two parameters at disposal will allow to consider limiting cases.

\subsection{Two-parameter family of Poisson-Lie dual models}\label{twoparam}

In what follows we  slightly modify  the current algebra \eqn{poissontaudefdrinf} by introducing another imaginary parameter, $\alpha$, so to have  $\mathfrak{su}(2)$ and $\mathfrak{sb}(2,\mathbb{C})$ generators on an equal footing. This will allow to  formulate Poisson-Lie duality as a phase space rotation within ${SL}(2,\C)(\R)$, namely  an $O(3,3)$ transformation, which exchanges configuration space coordinates, $K^i$ with momenta $S_j$. The introduction of the new  parameter will make it possible to perform not only the limit  ${SL}(2,\C)\stackrel{\tau\rightarrow 0}{\rightarrow }T^* SU(2)$ but also ${SL}(2,\C)\stackrel{\alpha\rightarrow 0}{\rightarrow }T^* SB(2,\C)$.

To this, let us introduce the two-parameter generalisation of the algebra \eqn{poissontaudefdrinf}   as follows:
\beqa\label{twoparamfamilyalgebra}
 \{S_i(\sigma), \, S_j(\sigma')\; \}&=&i\alpha{\epsilon_{ij}}^k S_k(\sigma)\delta(\sigma-\sigma')- \alpha^2 \hat C \delta_{ij}\delta'(\sigma-\sigma')  \nonumber\\
\{K^i(\sigma), K^j(\sigma') \}&=&i\tau{f^{ij}}_k K^k(\sigma)\delta(\sigma-\sigma')+\tau^2 \hat C h^{ij} \delta'(\sigma-\sigma')  \\ 
\{S_i(\sigma), K^j(\sigma')\;\}&=&\left[i\alpha{\epsilon_{ki}}^j K^k(\sigma)  +i\tau {f^{jk}}_i S_k(\sigma)  \right] \delta(\sigma-\sigma')  +( i \alpha\hat C'\delta_i^j - i \tau \hat C {\epsilon_i}^{j3})  \delta'(\sigma-\sigma').\nonumber
\eeqa
It is immediate to check that, in the limit $i\tau\rightarrow 0$, the latter reproduces the semi-direct sum $\mathfrak{su}(2)(\R) \dot\oplus \mathfrak{a}$, while the limit $ i \alpha\rightarrow 0$ yields  $\mathfrak{sb}(2,\C)(\R) \dot\oplus \mathfrak{a}$. For all non-zero values of the two parameters, the algebra is homomorphic to $\mathfrak{c}_2$, with central extensions. The central charges, $\hat C, \hat C'$ will be fixed in a while.

By direct calculation one easily verifies that, upon suitably rescaling the fields, one gets back the dynamics of the WZW model, if the Hamiltonian is deformed as follows:
\begin{equation}\label{directham}
\begin{aligned}
{} & H_{\tau, \alpha}=\lambda^2 \int_{\mathbb{R}} d\sigma \, S_I \left(\mathcal{M}_{\tau, \alpha} \right)^{IJ} S_J \\ & = \lambda^2 \int_{\mathbb{R}} d\sigma \left[S_i(\mathcal{M}_{\tau, \alpha})^{ij}  S_j+K^i(\mathcal{M}_{\tau, \alpha})_{ij}  K^j+S_i  {(\mathcal{M}_{\tau, \alpha})^i}_jK^j +K^i {(\mathcal{M}_{\tau, \alpha})_i}^jS_j \right],
\end{aligned}
\end{equation}
with 
$S_I=(S_i, K^i)$,  and 
\begin{equation}\label{Mmetdef}
\mathcal{M}_{\tau, \alpha}=\begin{pmatrix}
  \frac{1}{(i\alpha)^2}\left[(1+a^2 \bar \tau^4)\delta^{ij}-\bar\tau^2(1+a^2){\epsilon}^{ip3}\delta_{pq} \epsilon^{jq3} \right] &\left[ i\bar\tau(1+a^2){\epsilon}^{ip3}+a(1+\bar \tau^2)\delta^{ip}\right]\delta_{pj}\\
  \delta_{ip}\left[- i\bar \tau  (1+a^2)\epsilon^{pj3}+a(1+\bar \tau^2)\delta^{pj} \right] & (i\alpha)^2(1+a^2)\delta_{ij}
 \end{pmatrix} 
\end{equation}
where $i\bar\tau= i\tau\,i\alpha$.
Indeed, by rescaling the fields according to
\be
\bar S_j = \frac{S_j}{i\alpha},\;\;\; \bar K^j= i\alpha K^j 
\ee
the Hamiltonian for the fields $\bar S_j, \bar K^j$ takes the same form as  Eq. \eqn{hamsk} and 
the Poisson brackets  of the rescaled fields yield back   the algebra \eqn{poissontaudefdrinf} if the central charges are chosen as follows: 
\be
\hat C= \frac{a}{\lambda^2 \left(1-a^2 \bar \tau^2 \right)^2}, \;\;\; \hat C'=- \frac{1+a^2 \bar \tau^2 }{2\lambda^2\left(1-a^2\bar \tau^2  \right)^2}.
\ee

This is exactly what we were looking for, since the role of the $\mathfrak{su}(2)$ and $\mathfrak{sb}(2,\mathbb{C})$ generators is now completely symmetric. It is easy to check that the two-parameter model indeed reproduces  the original WZW dynamics for $i\tau \to 0$,  with Poisson algebra  $\mathfrak{c}_1=\mathfrak{su}(2)(\mathbb{R}) \dot{\oplus} \, \mathfrak{a} $. The  limit  $i\alpha \to 0$ limit yields the algebra  $\mathfrak{c}_3=\mathfrak{sb}(2, \mathbb{C})(\mathbb{R}) \dot{\oplus} \, \mathfrak{a} $ with central  extension, although the Hamiltonian appears to be singular in such a limit.  We shall come back to this issue later on. For all  other values of $\alpha$ and $\tau$ the algebra is isomorphic to $\mathfrak{c}_2 \simeq \mathfrak{sl}(2,\mathbb{C})(\mathbb{R})$, with $K^i, S_i$ respectively playing the role of configuration space coordinates and momenta. 

Since now the role of $S_i$ and $K^i$ is symmetric, if we exchange the momenta $S_i$ with the configuration space fields $K^i$  we obtain a new two-parameter family of models, with the same target phase space, but with the role of coordinates and momenta  inverted. The transformation
\begin{equation}
\tilde{K_i}(\sigma)=S_i(\sigma), \quad \tilde{S^i}(\sigma)= K^i(\sigma)
\end{equation}
is an $O(3,3)$ rotation in the target phase space $SL(2,\mathbb{C})$. 

Explicitly, under such a rotation we obtain the dual Hamiltonian
\begin{equation}\label{dualham}
\tilde{H}_{\tau, \alpha}=\lambda^2 \int_{\mathbb{R}} d\sigma \left[\tilde{K}_i(\mathcal{M}_{\tau, \alpha})^{ij}  \tilde{K}_j+ \tilde{S}^i (\mathcal{M}_{\tau, \alpha})_{ij} \tilde{S}^j+\tilde{K}_i {(\mathcal{M}_{\tau, \alpha})^i}_j\tilde{S}^j + \tilde{S}^i {(\mathcal{M}_{\tau, \alpha})_i}^j \tilde{K}_j\right],
\end{equation}
and  dual Poisson algebra
\beqa \label{dcalgebra}
 \{\tilde{K}_i(\sigma), \tilde{K}_j(\sigma') \}&=&i\alpha{\epsilon_{ij}}^k \tilde{K}_k(\sigma)\delta(\sigma-\sigma')- \alpha^2 \hat C \delta_{ij}\delta'(\sigma-\sigma') \nonumber\\
\{\tilde{S}^i(\sigma), \tilde{S}^j(\sigma')\} &=&i\tau{f^{ij}}_k \tilde{S}^k(\sigma)\delta(\sigma-\sigma') +\tau^2 \hat C h^{ij} \delta'(\sigma-\sigma') \\ 
\{\tilde{K}_i(\sigma), \tilde{S}^j(\sigma')\}&=&\left[i\alpha{\epsilon_{ki}}^j \tilde{S}^k(\sigma)  +i\tau {f^{jk}}_i \tilde{K}_k(\sigma)  \right] \delta(\sigma-\sigma') +  ( i \alpha\hat C'\delta_i^j - i \tau \hat C {\epsilon_i}^{j3})  \delta'(\sigma-\sigma')\nonumber
\eeqa
which makes it clear that this new two-parameter family of models has target configuration space the group manifold of $SB(2,\mathbb{C})$, spanned by the fields $\tilde{K}_i$, while momenta $\tilde{S}^i$ span the fibres of the target phase space. Hence, this represents by construction a family of dual models.

Note, however, that the limit $i\alpha \to 0$, although giving a well-defined Poisson algebra as a semi-direct sum, does not bring to a  well-defined dynamics on $T^*SB(2,\C)$, since the Hamiltonian becomes singular. As we shall see in the next section, this seems to be related to the impossibility of obtaining the family of dual Hamiltonians \eqn{dualham} from a continuous deformation of a Hamiltonian WZW model on the cotangent space $T^*SB(2,\C)$. This obstruction has a topological explanation  in the simple fact that $T^*SB(2,\C)$, differently from $T^*SU(2)$, is not homeomorphic to $SL(2,\C)$.   In the next section we shall introduce a WZW model with $SB(2,\C)$ as configuration space, in the Lagrangian approach, and shall look for a Hamiltonian description by means of  canonical Legendre transform. We shall see that, in order to make contact 
with  one of the dual models described by the two-parameter family \eqn{dualham},  we need to introduce a true deformation of the dynamics, a topological modification of the phase space and extra terms in the Hamiltonian. 

Going back to the Hamiltonian \eqn{dualham} we want to show here that, although it has not been obtained from an action principle, nevertheless it is possible to exhibit an action from which it can be derived.  Following the standard approach of \cite{loop, R89},  we shall  write the action in the first order formalism. To this,  two ingredients are needed: the symplectic form responsible for the  current algebra  (\ref{dcalgebra}) and the Hamiltonian  (\ref{dualham}) expressed  in terms of the original fields $g \in SU(2)(\mathbb{R})$ and $\ell \in SB(2,\mathbb{C})(\mathbb{R})$. The target phase space $\Gamma_2$ can be identified with  $SU(2)(\mathbb{R}) \times SB(2,\mathbb{C})(\mathbb{R})$ as a manifold, and we define 
\begin{equation}
-i\alpha \hat{C}g^{-1}\partial_{\sigma} g=i\delta^{kp}\tilde{K}_p e_k, \quad i\tau\hat{C} \ell^{-1}\partial_{\sigma} \ell=i(h^{-1})_{kp} \tilde{S}^p \hat{e}^k,
\end{equation}
with $\hat{e}^k$ the generators of the $\mathfrak{sb}(2,\mathbb{C})$ algebra (see App. \ref{sl2c}). It can be shown that the symplectic form which yields the  current algebra (\ref{dcalgebra}) is the following (see Appendix \ref{sympl} for more details about this construction):
\begin{equation}\label{symplecticformgl}
\begin{aligned}
\omega=  {} & \alpha^2 \hat{C}\int_{\mathbb{R}} d\sigma \, \text{Tr}_{\mathcal{H}}\left[g^{-1}dg \wedge \partial_{\sigma} (g^{-1}dg) \right]-\tau^2 \hat{C}\int_{\mathbb{R}} d\sigma \, \text{Tr}_{\mathcal{H}}\left[\ell^{-1}d\ell \wedge \partial_{\sigma} (\ell^{-1}d\ell) \right] \\ &+i\bar{\tau} \hat{C}\int_{\mathbb{R}} d\sigma \,  \text{Tr}_{\mathcal{H}}\left\{[g^{-1}dg, \ell^{-1}\partial_{\sigma} \ell] \wedge (\ell^{-1}d\ell-(\ell^{-1}d\ell)^{\dagger})\right\} \\ & +i\bar{\tau} \hat{C}\int_{\mathbb{R}} d\sigma \,  \text{Tr}_{\mathcal{H}}\left\{ [\ell^{-1}d\ell, g^{-1}\partial_{\sigma} g] \wedge (g^{-1}dg-(g^{-1}dg)^{\dagger})  \right\} \\ & +i\alpha \hat{C}' \int_{\mathbb{R}^2} d\sigma \, d\sigma' \partial_{\sigma} \delta(\sigma-\sigma') \left\{ \text{Tr}_{\mathcal{\eta}} \left[g^{-1}dg(\sigma) \wedge \ell^{-1}d\ell(\sigma')\right]\right\} \\ & -i\tau \hat{C} \int_{\mathbb{R}^2} d\sigma \, d\sigma' \partial_{\sigma} \delta(\sigma-\sigma') \left\{\text{Tr}_{\mathcal{H}} \left[g^{-1}dg(\sigma) \wedge \ell^{-1}d\ell(\sigma') \right] \right\}.
\end{aligned}
\end{equation}
The products denoted by $\text{Tr}_{\mathcal{H}}$ and $\text{Tr}_{\eta}$ are the two $SL(2,\mathbb{C})$ products defined in \eqn{riemannianmetric} and \eqn{o33metric} respectively.
The symplectic form is not closed, therefore an  action in the first order formalism may be defined according to 
\begin{equation} \label{foaction}
S_2=\int \omega-\int H_{|g,\ell} dt,
\end{equation}
where $\omega$ has to be integrated on a two-surface and $H_{|g,\ell}$ denotes the Hamiltonian expressed in terms of the original fields $g$ and $\ell$. 
When the symplectic form is exact,  the   surface integral  reduces to the standard  integration of the canonical Lagrangian 1-form along the boundary of the surface. However, this is not the case for our symplectic form and some care is needed. Here one has to consider the closed curve $\gamma$ on $\Gamma_2$, described by functions $g(t, \sigma) : \mathbb{R} \times S^1 \to SU(2)$ and $\ell(t, \sigma) : \mathbb{R} \times S^1 \to SB(2, \mathbb{C})$, as well as the surface $\tilde{\gamma}$ of which it is the boundary: $\partial \tilde{\gamma}=\gamma$. The surface $\tilde{\gamma}$ can then be described by extensions of $g$ and $\ell$ defined such that $\tilde{g}(t, \sigma,y=1)=g(t, \sigma)$, $\tilde{\ell}(t,x, y=1)=\ell(t, \sigma)$ and $\tilde{g}(t, \sigma,y=0)=\tilde{\ell}(t,\sigma,y=0)=1$, mimicking the Wess-Zumino term construction on a $3$-manifold with fields extended from the source space. The action can then be written explicitly with (\ref{foaction}). Note also that the integration of the first two terms in (\ref{symplecticformgl}) following this recipe results in two WZ terms. The construction is totally symmetric with respect to the exchange of momenta with configuration fields, therefore, the same construction furnishes an action principle for the Hamiltonian \eqn{directham}.

Summarising, we reformulated the WZW model on $SU(2)$ within an alternative canonical picture based on a two-parameter deformation of the current algebra and the Hamiltonian, in which the role of momenta  and configuration space fields  is made symmetric. By sending to zero either parameter we recover the  original current algebra structure $\mathfrak{su}(2)(\mathbb{R}) \dot \oplus \mathfrak{a}(\mathbb{R})$ or the natural dual one $\mathfrak{sb}(2,\mathbb{C})(\mathbb{R})\dot\oplus \mathfrak{a}(\mathbb{R})$. By performing an $O(3,3)$ transformation over $SL(2,\mathbb{C})$, which is the deformed target phase space of the system, 
we obtain a two-parameter family of Hamiltonian models with target configuration space $SB(2,\mathbb{C})$, which represents, by construction, the Poisson-Lie dual family of the $SU(2)$ family we started with. 

As a further goal, in parallel to what is done for the $SU(2)$ family, where the limit $i\tau\rightarrow 0$ yields back the semi-Abelian model with target phase space $T^*SU(2)$, we would like to perform the limit $i\alpha\rightarrow 0$  to recover the semi-Abelian dual model with target phase space $T^*SB(2,\C)$. We have seen that, while the current algebra is well-defined in such a limit, yielding the semi-direct sum $\mathfrak{sb}(2,\C)(\R)\dot\oplus \mathfrak{a}$, the Hamiltonian is singular.  We have argued that  this may be related to the different topology of phase spaces $SL(2,\mathbb{C})$ and $T^* SB(2,\mathbb{C})$. This issue will be  addressed at the end of Sect. \ref{hamiltonianformsb}.

\section{Lagrangian WZW model on $SB(2,\mathbb{C})$}
\label{sectwzwsb2c}

In the previous section we have obtained a whole family of dual models having $SB(2,\mathbb{C})$ as target configuration space, which makes it meaningful to look for a dual model having the tangent bundle of $SB(2,\mathbb{C})$ as carrier space from the beginning. However,  the latter group algebra is not semi-simple, which means that the Cartan-Killing metric is degenerate. The problem of constructing a WZW model for non-semisimple groups is not new - see for example \cite{nappi}, where the 2-d Poincar\'e group is considered. In our case,  it does not seem to be possible to define any bilinear pairing on $\mathfrak{sb}(2,\mathbb{C})$ which is both non-degenerate and bi-invariant at the same time. As in \cite{marottapcm},  we could  use 
  the induced metric (\ref{sbmetric}), $h^{ij}= \delta^{ij}+\epsilon^{i\ell 3}\delta_{\ell k}\epsilon^{jk3}$,  which is obtained from  restricting the Riemannian metric \eqn{sb2cproduct} of $\mathfrak{sl}(2,\C)$. This  is non-degenerate and positive-definite, and 
  only invariant under left $SB(2,\mathbb{C})$ action. 
A natural WZW action would then be:
\begin{equation} \label{sbaction}
S_1=\frac{1}{n\lambda^2}\int_{\Sigma} \mathcal{T}r\left[\phi^*(\ell^{-1}d\ell) \stackrel{\wedge}{, }* \phi^*(\ell^{-1}d \ell) \right]+\frac{\kappa}{m \pi}\int_{\mathcal{B}}\mathcal{T}r\left[\tilde{\phi}^*\left(\tilde{\ell}^{-1}d \tilde{\ell} \stackrel{\wedge}{, } \tilde{\ell}^{-1}d \tilde{\ell} \wedge \tilde{\ell}^{-1}d \tilde{\ell} \right) \right],
\end{equation}
with $\phi: \Sigma \ni (t,\sigma) \rightarrow \ell \in SB(2,\mathbb{C})$,    while $\tilde{\phi}$ 
and $\tilde{\ell}$ are the related extensions to $\mathcal{B}$, and $\mathcal{T}r :=((,))$ as in \eqn{sb2cproduct}. 
Following our discussion, the so-defined Lagrangian is left and right invariant under $SU(2)$ action and only left-invariant under $SB(2,\mathbb{C})$ action. 
However, it is immediate to check that the WZ term is identically zero for such a product. Indeed, on introducing the notation  
\be
B_\mu=\ell^{-1}\del_\mu  \ell=  B_{\mu j } \hat e^j,
\ee
   with $\hat e^j$ indicating the generators of $\mathfrak{sb}(2,\C)$ 
    and $\tilde B_\mu$ the extension to $\mathcal{B}$, we have
\begin{equation*}
\begin{aligned}
\left( \left( \tilde{\phi}^*\left[\tilde{\ell}^{-1}d \tilde{\ell} \stackrel{\wedge}{, } \tilde{\ell}^{-1}d \tilde{\ell} \wedge \tilde{\ell}^{-1}d \tilde{\ell} \right] \right)\right) & =-i \, \dd^3 y \, \epsilon^{\alpha \beta \gamma} \tilde{B}_{\alpha i} \tilde{B}_{\beta j} \tilde{B}_{\gamma k} h^{kp} {f^{ij}}_p \\
&=  -2i\,\dd^3 y \,\epsilon^{\alpha \beta \gamma}\left(\tilde{B}_{\alpha i} \tilde{B}_{\beta j} \tilde{B}_{\gamma 1} \epsilon^{ij2}-\tilde{B}_{\alpha i} \tilde{B}_{\beta j} \tilde{B}_{2 \gamma} \epsilon^{ij1}\right),
\end{aligned}
\end{equation*}
which is obviously vanishing. This means that  $h^{ij}$ is not a viable product  to define a WZW model on $SB(2,\mathbb{C})$. 

Our proposal is then  to use the Hermitian product $h_N$ defined in Eq. (\ref{newproduct}). The action of the model will be given by Eq. \eqn{sbaction} with ${\mathcal T}r(u,v)\rightarrow \Tr (u^\dag v)$ and  $n,m,$ integer coefficients to be  determined later. 
In terms of the latter, the WZ term can be checked to be non-zero and consistent with the equations of motion one expects to obtain. Indeed,  on separating the diagonal and off-diagonal part of the product, as 
\be\label{hN}
h_N^{ij}=\frac{1}{2} h^{ij}+a^{ij}, 
\ee
the only contribution to the volume integral  in \eqn{sbaction} comes from the off-diagonal term, $a^{ij}$,  since we just showed that the WZ term vanishes with the metric $h^{ij}$. We have 
\beqa
\int_{\mathcal{B}} \dd^3 y \, \epsilon^{\alpha \beta \gamma} \tilde{B}_{\alpha i} \tilde{B}_{\beta j} \tilde{B}_{\gamma k} h_N^{kp} {f^{ij}}_p&=&\frac{1}{2}\int_{\mathcal{B}} \dd^3 y \, \epsilon^{\alpha \beta \gamma} \tilde{B}_{\alpha i} \tilde{B}_{\beta j} {\tilde B}_{\gamma k}  a^{kp} {f^{ij}}_p \nonumber\\
&=&\frac{i}{2}\, \int_{\mathcal{B}} \dd^3 y \, \epsilon^{\alpha \beta \gamma} \epsilon^{ijk} \tilde{B}_{\alpha i} \tilde{B}_{\beta j} \tilde{B}_{\gamma k}.
\eeqa
The latter has the same form as the WZ term on the $SU(2)$ group manifold, which means that the variation is formally  the same, leading to the same contribution to the equations of motion but now with $\mathfrak{sb}(2,\mathbb{C})$-valued currents:
\begin{equation}
\delta S_{1,WZ}=\frac{1 }{m\pi} \int_{\Sigma} d^2\sigma \, \epsilon^{ijk} \, B_{0 j} B_{1k} \left( \ell^{-1}\delta \ell \right)_i
\end{equation}
with $B_0=\ell^{-1}\del_t \ell, \, B_1=\ell^{-1}\del_\sigma \ell$. 

As for the quadratic term in the action \eqn{sbaction}, on using Eq. \eqn{hN} we have 
 \be
 \int_{\Sigma} \dd^2 \sigma \, \mathcal{T}r \left[\phi^*(\ell^{-1}d\ell) \stackrel{\wedge}{, }* \phi^*(\ell^{-1}d \ell) \right]= \int_{\Sigma} \dd^2 \sigma \, B_{\mu i}B^\mu_j \Tr{\hat e}^{i \dag} \hat e^ j 
= \frac{1}{2} \int_{\Sigma} \dd^2 \sigma \, ((B_{\mu }, B^\mu))
 \ee
because the off-diagonal contribution proportional to $a^{ij}$ vanishes. Therefore, in absence of the WZ term, the two products yield the same result, up to a numerical factor,  and agree with previous findings for the PCM \cite{marottapcm}. 
For the variation of this term with respect to small variations of $\ell$ we will need the following relation:
\begin{equation}
\label{deltai}
\delta B_{\mu}=-\ell^{-1}\delta \ell B_{\mu}+\ell^{-1}\partial_{\mu} \delta \ell,
\end{equation}
\beqa
 \delta(({B_{\mu}}, B^{\mu} )) &=&((\delta B_\mu, B^\mu))+ ((B_\mu, \delta B^\mu))\nonumber\\
 &=& 2 \left[ ((B^\mu, [B_\mu, \ell^{-1} \delta \ell])) - ((\del_\mu B^\mu, \ell^{-1}\delta l)) + \del_\mu ((B^\mu, \ell^{-1}\delta \ell))\right]
\eeqa
and after integration  one obtains
\begin{equation*}
\delta S_{1, \text{quad}}=\frac{1}{n\lambda^2} \int_{\Sigma} d^2 \sigma \left[((-\partial^{\mu} B_{\mu},\ell^{-1} \delta \ell))+((B^{\mu},[B_{\mu},\ell^{-1}\delta \ell])) \right].
\end{equation*}
In order to make the comparison with the $SU(2)$ model more transparent we fix $n=4$ and introduce the notation  $\hat A_i = B_{0,i}, \hat J_i= B_{1 i}$. 
We have then 
\begin{equation}
\label{eomkinsb}
\delta S_{1, \text{quad}}=-\frac{1}{4\lambda^2}\int_{\Sigma} d^2\sigma \left[h^{ij}\left(\partial_t \hat A_j-\partial_{\sigma}\hat J_j \right)-{f^{pi}}_q h^{qj}\left(\hat A_p \hat A_j-\hat J_p \hat J_j \right) \right](\ell^{-1}\delta \ell)_i.
\end{equation}
By collecting all terms, the resulting equations of motion can then be written as follows:
\begin{equation}
h^{ij}\left(\partial_t \hat A_j-\partial_{\sigma}\hat J_j \right)-{f^{pi}}_q h^{qj}\left(\hat A_p \hat A_j-\hat J_p \hat J_j \right)=-\frac{4\kappa \lambda^2}{m\pi} \epsilon^{ipj} \, \hat A_p \hat J_j,
\end{equation}
and we also have the usual integrability condition coming from the Maurer-Cartan equation for the Maurer-Cartan one-forms $\ell^{-1}d \ell$:
\begin{equation}
\partial_t \hat J-\partial_{\sigma}\hat A=-[\hat A,\hat J].
\end{equation}
Looking at the equations of motion, by analogy with the $SU(2)$ case we will fix $m=24$.

\subsection{Spacetime geometry}
The Lagrangian model which  has been derived  in the previous section  furnishes a possible  spacetime background on which strings propagate. Topologically it is the manifold of the group $SB(2,\C)$, a noncompact manifold, which can be embedded in $\R^4$ by means of the following parametrization
\be
  \ell= y_0 \mathds{1}_2 + 2i y_i \hat e^i 
  \ee
   with $y_\mu, \mu=0,\dots,3$ global real coordinates,  $\tilde e^i$  the generators of the group (see def. \eqn{sbgen}) and the constraint $y_0^2-y_3^2=1$.  Its geometry is characterised by a  metric tensor and a $\mathds{B}$-field, which are easier to compute in terms of  a local parametrisation.  
We first  rewrite the  action as  
\begin{equation}\label{sbactionf}
S_1=\frac{1}{4\lambda^2}\int_{\Sigma} d^2 \sigma {B}_{\mu i}{B}^{\mu}_j h^{ij}+\frac{\kappa}{24 \pi}\int_{\mathcal{B}} d^3 y \, \epsilon^{\alpha \beta \gamma} \tilde{B}_{\alpha i} \tilde{B}_{\beta j} \tilde{B}_{\gamma k}h_N^{kp} {f^{ij}}_p.
\end{equation}
with   $i B_i \hat e^i = \ell^{-1} d \ell$ the Maurer-Cartan one-form on the group. 
We  thus parametrise a generic element $\ell \in SB(2,\mathbb{C})$ according to 
\begin{equation}
\ell=\begin{pmatrix}
  \chi & \psi e^{i\theta}\\
  0 & \frac{1}{\chi}
 \end{pmatrix},
\end{equation}
with $\chi, \psi\in \R, \chi> 0$ and $ \theta \in (0,2\pi)$. In this way we can write 
\begin{equation}
\ell^{-1} d \ell=\begin{pmatrix}
  \frac{1}{\chi} d \chi & \frac{1}{\chi} e^{i\theta} d\psi+i\frac{\psi}{\chi}e^{i\theta}d\theta+\frac{\psi}{\chi^2}e^{i\theta} d\chi \\
  0 & -\frac{1}{\chi} d\chi
 \end{pmatrix}.
\end{equation}
Since the generators of the $\mathfrak{sb}(2,\mathbb{C})$ algebra can be written as
\begin{equation}
\hat{e}^k=\frac{1}{2}\delta^{ki}\left(i \sigma_i+\epsilon^k_{i3} \sigma_k \right),
\end{equation}
or, explicitly,
\begin{equation}
\hat{e}^1=\begin{pmatrix}
  0 & i\\
  0 & 0
 \end{pmatrix}, \quad \hat{e}^2= \begin{pmatrix}
  0 & 1\\
  0 & 0
 \end{pmatrix}, \quad \hat{e}^3=\frac{i}{2}\begin{pmatrix}
  1 & 0\\
  0 & -1
 \end{pmatrix},
\end{equation}
the components of the Maurer-Cartan one-form $\ell^{-1} d \ell$ have the form:
\beqa
B_1&=&-\frac{\psi}{\chi^2}\cos\theta d\chi-\frac{1}{\chi}\cos \theta d\psi +\frac{\psi}{\chi}\sin\theta d\theta, \\
B_2&=&\frac{\psi}{\chi^2}\sin \theta d\chi+\frac{\sin \theta }{\chi}d\psi +\frac{\psi}{\chi}\cos \theta d\theta,\\
B_3&=&-\frac{2}{\chi}d\chi.
\eeqa
On using the explicit expression  of the product $h^{ij}$ the quadratic  term of the model yields then
\begin{equation}
S_{1 quad}= \frac{1}{4\lambda^2}\int_{\Sigma} d^2\sigma \left[\left(\frac{\psi^2}{2\chi^4}+\frac{4}{\chi^2} \right) \partial_{\mu}\chi \, \partial^{\mu} \chi+\frac{1}{2\chi^2} \partial_{\mu} \psi \, \partial^{\mu} \psi +\frac{\psi^2}{2\chi^2} \partial_{\mu}\theta \, \partial^{\mu} \theta+\frac{\psi}{\chi^3} \partial_{\mu} \chi \, \partial^{\mu} \psi\right].
\end{equation}
Analogously, the WZ term can be calculated in local coordinates to give:
\begin{equation*}
\epsilon^{\alpha \beta \gamma} \tilde{B}_{\alpha i} \tilde{B}_{\beta j} \tilde{B}_{\gamma k} h^{kp}_N {f^{ij}}_p=2\epsilon^{\alpha \beta \gamma} \tilde{B}_{\alpha 1} \tilde{B}_{\beta 2} \tilde{B}_{\gamma 3}=4 \epsilon^{\alpha \beta \gamma} \frac{\tilde{\psi}}{\tilde{\chi}^3} \partial_{\alpha}\tilde{\chi} \partial_{\beta} \tilde{\psi} \partial_{\gamma} \tilde{\theta}=-2 \epsilon^{\alpha \beta \gamma} \partial_{\alpha}\left(\frac{\tilde{\psi}}{\tilde{\chi}^2} \partial_{\beta} \tilde{\psi} \partial_{\gamma}\tilde{\theta} \right).
\end{equation*}
Hence, by means of  Stokes theorem on the latter contribution, the total  action can be rewritten as
\begin{equation}
\begin{aligned}
S_1=  \frac{1}{4 \lambda^2} {} & \int_{\Sigma} d^2 \sigma \bigg[ \left(\frac{\psi^2}{2\chi^4}+\frac{4}{\chi^2} \right) \partial_{\mu}\chi \, \partial^{\mu} \chi+\frac{1}{2\chi^2} \partial_{\mu} \psi \, \partial^{\mu} \psi +\frac{\psi^2}{2\chi^2} \partial_{\mu}\theta \, \partial^{\mu} \theta+\frac{\psi}{\chi^3} \partial_{\mu} \chi \, \partial^{\mu} \psi\\ & - \frac{\kappa \lambda^2}{3 \pi}\frac{\psi}{\chi^2} \epsilon^{\mu\nu} \partial_{\mu} \psi \, \partial_{\nu} \theta \bigg].
\end{aligned}
\end{equation}
By identifying the latter  with the Polyakov action
\begin{equation}
\int d^2 \sigma \left(G_{ij} \partial_{\alpha} X^i \partial^{\alpha}X^j + B_{ij} \epsilon^{\alpha \beta}\partial_{\alpha} X^i \partial_{\beta} X^j \right),
\end{equation}
with $X^i \equiv \left(\chi, \psi, \theta \right)$, the background spacetime metric and $B$-field read 
\begin{equation}\label{metgeo}
G_{ij}=\frac{1}{4\lambda^2}\begin{pmatrix}
 \left(\frac{\psi^2}{2\chi^4}+\frac{4}{\chi^2} \right) & \frac{\psi}{{ 2\chi^3}} & 0 &\\
\frac{\psi}{{2 \chi^3}} &\frac{1}{2\chi^2} & 0 \\
  0& 0 &{\frac{\psi^2}{2\chi^2}}
 \end{pmatrix}, \quad B_{\psi \theta}= - \frac{\kappa}{12\pi} \frac{\psi}{ \chi^2}.
\end{equation}
Hence, the spacetime background is a non-compact 3-d Riemannian manifold, embedded in $\R^4$ with the topology of the group manifold of $SB(2,\C)$ and its geometry is described by the following above metric and antisymmetric $B$-field.
The $B$-field is not closed, thus yielding a 
3-form $\mathds{H}$-flux. 

It may be useful  to express the metric and the $B$-field in terms of global coordinates in $\R^4$. It can be easily checked that the embedding map reads
\be 
\psi= 2\sqrt{y_1^2+y^2_2},\;\;\; \chi= y_0-y_3\;\;\; \theta= -\arctan \frac{y_2}{y_1}.
\ee
Then the metric  in \eqn{metgeo} is obtained by the following Lorentzian metric in $\R^4$
\beqa
\mathds{G}_4&=&\frac{2}{{(y_0-y_3)^2}} \Bigl[ f_a(-dy_0\otimes dy_0+ dy_3\otimes dy_3)+ f_b (dy_1\otimes dy_1+ dy_2\otimes dy_2)\Bigr. \nonumber\\
&+& \Bigl.f_c d(y_0-y_3)\otimes (y_1dy_1+y_2dy_2)\Bigr]
\eeqa
with 
\be f_a= {y_1^2+y_2^2+ 2(y_0-y_3)^2 },\;\;\; f_b=1,\;\,\;f_c= \frac{2}{(y_0-y_3)}.
\ee
Upon imposing the constraint $(y_0-y_3)(y_0+y_3)=1$, which characterises the submanifold, we get:
\beqa
\mathds{G}_3&=&\frac{2}{{(y_0-y_3)^2}} \Bigl[ \frac{f_a}{(y_0-y_e)^2} d(y_0-y_3) \otimes d(y_0-y_3)+ f_b (dy_1\otimes dy_1+ dy_2\otimes dy_2) \Bigr.\nonumber\\
&+&\Bigl. f_c d(y_0-y_3)\otimes (y_1dy_1+y_2dy_2)\Bigr].
\eeqa
Analogously, we may write the two-form $B$ in terms of  global $\R^4$ coordinates. We obtain
\be
B= \frac{ \kappa}{3 \pi} \frac{1}{(y_0-y_3)^2}{dy_1\wedge dy_2}
\ee
with $\mathds{H}$-flux 
\be
\mathds{H}= d B =  -\frac{2 \kappa}{3 \pi} \frac{1}{(y_0-y_3)^3}d(y_0-y_3)\wedge {dy_1\wedge dy_2}.
\ee

\subsection{Dual Hamiltonian formulation}
\label{hamiltonianformsb}

Let us consider first the situation in which the WZ term is missing ($\kappa=0$). In this case the equations of motion have the simpler form
\begin{equation}
h^{ij}\left(\partial_t \hat A_j-\partial_{\sigma}\hat J_j \right)={f^{pi}}_q h^{qj}\left(\hat A_p \hat A_j-\hat J_p \hat J_j \right),
\end{equation}
\begin{equation}
\partial_t \hat J-\partial_{\sigma} \hat A=-[\hat A,\hat J]
\end{equation}
and we have a clear Lagrangian picture. In particular, we are able to define the left momenta 
\begin{equation}
\hat I^i=\frac{\delta \mathscr{L}_1}{\delta \hat A_i}=\frac{1}{2 \lambda^2} \hat A_j h^{ij}
\end{equation}
which can be inverted for the generalized velocities to write the Hamiltonian:
\begin{equation*}
H_1=\frac{1}{4 \lambda^2}\int_{\mathbb{R}} d\sigma \left(\hat I^i \hat I^j h_{ij}+\hat J_i \hat J_j h^{ij} \right)
\end{equation*}
and analogously to the $SU(2)$ case, the pair $(\hat A, \hat J)$ identifies the cotangent bundle of $SB(2,\mathbb{C})$, with $\hat I$ fibre coordinates. 

Following the usual approach we can then obtain the equal-time Poisson brackets from the action functional:
\begin{equation}
\begin{aligned}
{} & \{\hat I^i (\sigma), \hat I^j(\sigma')\}=2\lambda^2{f^{ij}}_k \hat I^k(\sigma)\delta(\sigma-\sigma') \\ &
\{\hat I^i (\sigma), \hat J_j(\sigma')\}=2\lambda^2\left[{f^{ki}}_j \hat J_k(\sigma)\delta(\sigma-\sigma')-\delta_j^i \delta'(\sigma-\sigma') \right] \\ &
\{\hat J_i(\sigma),\hat J_j(\sigma') \}=0,
\end{aligned}
\end{equation}
from which, together with the Hamiltonian $H_1$, the equations of motion follow:
\begin{equation}
\partial_t \hat{I}^k(\sigma)=h^{ij} {\delta^k}_{j} \, \partial_{\sigma}\hat{J}_i(\sigma)+h_{ij}{f^{jk}}_p \, \hat{I}^i(\sigma) \hat{I}^p(\sigma)+h^{ij}{f^{kp}}_j  \, \hat{J}_i(\sigma) \hat{J}_p (\sigma),
\end{equation}
\begin{equation}
\partial_t \hat{J}_k(\sigma)=h_{ij}\left[{f^{pi}}_k \, \hat{I}^j(\sigma)\hat{J}_p (\sigma)+{\delta^i}_k \, \partial_{\sigma}\hat{I}^j(\sigma) \right].
\end{equation}
If we now introduce the WZ term contribution, the equations of motion get modified and a new term appears:
\begin{equation}
\label{eqmsbmoment}
\begin{aligned}
\partial_t \hat{I}^k(\sigma) {}  =& h^{ik} \, \partial_{\sigma}\hat{J}_i(\sigma)+h_{ij}{f^{jk}}_p \, \hat{I}^i(\sigma)\hat{I}^p(\sigma)+h^{ij}{f^{kp}}_j  \, \hat{J}_i(\sigma) \hat{J}_p (\sigma) \\ &-\frac{\kappa \lambda^2}{3 \pi}  h_{pi} \epsilon^{kpj} \, \hat{I}^i (\sigma) \hat{J}_j (\sigma),
\end{aligned}
\end{equation}
while the integrability condition does not receive any modification, as it should.

Assuming that the Hamiltonian remains the same after the inclusion of the WZ term, as it is the case for the $SU(2)$ model, we have to find the corresponding Poisson structure leading to the modified equations of motion. By inspection, Hamilton equations for momenta, obtained from the bracket $\{H_1, \hat{I}\}$, only involve the bracket  $\{\hat{I}, \hat{I} \}$, therefore we shall just modify the latter, by including a term proportional to $\hat{J}$.

It is straightforward to check that to obtain the right correction to  the equations of motion we have to modify the Poisson  brackets as follows:
\begin{equation}
\{\hat I^i (\sigma), \hat I^j(\sigma')\}=2\lambda^2\left[{f^{ij}}_k \hat I^k(\sigma)\delta(\sigma-\sigma')-w \, \epsilon^{ijp} \hat J_p (\sigma) \delta(\sigma-\sigma')\right],
\end{equation}
and the coefficient $w$ can be determined by direct comparison of Hamilton equation for $I$ with  (\ref{eqmsbmoment}). We find:
\begin{equation*}
w=\frac{2\kappa \lambda^2}{6 \pi}.
\end{equation*}
To summarise, the dynamics of the WZW model on $SB(2,\mathbb{C})$ is described by the following Hamiltonian
\begin{equation}
H_1= \frac{1}{4 \lambda^2}\int_{\mathbb{R}} d\sigma \left(\hat I^i \hat I^j h_{ij}+\hat J_i \hat J_j h^{ij} \right).
\end{equation}
 and Poisson algebra
\begin{equation}\label{PBd}
\begin{aligned}
{} & \{\hat I^i (\sigma), \hat I^j(\sigma')\}=2\lambda^2\left[{f^{ij}}_k \hat I^k(\sigma)\delta(\sigma-\sigma')-\frac{2\kappa \lambda^2}{6 \pi} \, \epsilon^{ijp} \hat J_p (\sigma) \delta(\sigma-\sigma')\right] \\ &
\{\hat I^i (\sigma), \hat J_j(\sigma')\}=2\lambda^2\left[{f^{ki}}_j \hat J_k(\sigma)\delta(\sigma-\sigma')-{\delta_j}^i \delta'(\sigma-\sigma') \right] \\ &
\{\hat J_i(\sigma), \hat J_j(\sigma') \}=0.
\end{aligned}
\end{equation}
This  is the semi-direct sum of an Abelian algebra and a Kac-Moody algebra associated to $SB(2, \mathbb{C})$, with a central extension,  just as expected.  
\\
Indeed, on defining 
\be
\hat S^i= \hat I^i-\frac{w}{2}\epsilon^{ij 3}\hat J_j
\ee
it is immediate to check that
\beqa
\{\hat S^i(\sigma), \hat S^j(\sigma')\}&=& 2\lambda^2 \left[{f^{ij}}_k \hat S^k(\sigma) \delta(\sigma-\sigma')- w\epsilon^{ij3}\delta'(\sigma-\sigma')\right]\\
\{\hat S^i(\sigma), \hat J_j(\sigma')\}&=& 2\lambda^2 \left[ {f^{ki}}_j \hat J_k(\sigma)  -\delta^{i}_j \delta'(\sigma-\sigma')\right].
\eeqa

We have already obtained   the same kind of algebra  as the $i \alpha \to 0$ limit of the algebra (\ref{twoparamfamilyalgebra}). 
The Hamiltonian however is not recovered as a limit of the dual family \eqn{dualham}, and we have already commented that this should be expected, because  $T^*SB(2,\mathbb{C})$, phase space of the former,   cannot be obtained by continuous deformation of $SL(2,\mathbb{C})$, phase space of the latter.

There is however another possibility, suggested by the form of the metric \eqn{Mmetdef}. 
Within the Hamiltonian picture, we have the freedom of defining another model, taking advantage of the fact the  Lie algebra $\mathfrak{sl}(2,\C)$ has another  metric structure, given by \eqn{o33metric}, which is  $O(3,3)$ invariant and non-degenerate. This provides a well defined  metric for  the Lie algebra $\mathfrak{sb}(2,\C)\dot\oplus \R^3$ as well. Therefore, we may declare the currents $\hat J$ to be valued in the Lie algebra $\mathfrak{sb}(2,\C)$, $\hat J= \hat J_i \hat e^i$, while the momenta $I$ to be valued in the Abelian algebra $\R^3$, $\hat I= \hat I^i \hat t_i$.   The metric \eqn{o33metric} will give 
\be
(\hat I, \hat I)= (\hat J, \hat J)= 0 , \;\;\; (\hat I, \hat J)= (\hat J, \hat I)= \hat I^i \hat J_j {\delta_i}^j
\ee
Thus, upon introducing the double field notation ${\bf \hat I} = (\hat I, \hat J)$, the Hamiltonian will be
\be
H_2= \zeta \int_{\mathbb{R}} \dd\sigma\, ({\bf \hat I}, {\bf \hat I} )= 2 \zeta\int_{\mathbb{R}} \dd\sigma\, \hat I^i \hat J_i.
\ee
The latter may be obtained from the two-parameter Hamiltonian of the dual family \eqn{dualham} in two steps.  We first introduce  a new Hamiltonian, which is a deformation of \eqn{dualham}, as
\be
\tilde{H}_{\text{def}}=\tilde{H}_{\tau, \alpha}-\lambda^2 \int_{\mathbb{R}} d\sigma \tilde{K}_i(\mathcal{M}_{\tau, \alpha})^{ij}  \tilde{K}_j
\ee
Then, we perform the limit $i\alpha\rightarrow 0$. This yields the wanted result if we suitably choose the parameter as $\zeta=a\lambda^2$:
\be\label{h2}
H_2=\lim_{i\alpha\rightarrow 0}  \tilde{H}_{\text{def}}.
\ee
On using  for  the Poisson algebra $\mathfrak{sb}(2,\mathbb{C})  \dot\oplus \mathbb{R}^3$ the brackets in (\ref{twoparamfamilyalgebra}) in the limit $i\alpha \rightarrow 0$:
\begin{equation}\label{PBdalphazero}
\begin{aligned}
{} & \{\tilde S^i (\sigma), \tilde S^j(\sigma')\}=i\tau{f^{ij}}_k \tilde S^k(\sigma)\delta(\sigma-\sigma')+\frac{a\tau^2}{\lambda^2} h^{ij}\delta'(\sigma-\sigma') \\ &
\{\tilde K_i (\sigma), \tilde S^j(\sigma')\}=i\tau{f^{jk}}_i \tilde K_k(\sigma)\delta(\sigma-\sigma')-\frac{i\tau a}{\lambda^2} {\epsilon_i}^{j3} \delta'(\sigma-\sigma') \\ &
\{\tilde K_i(\sigma), \tilde K_j(\sigma') \}=0,
\end{aligned}
\end{equation}
after identifying the currents as $\tilde S = \hat I$ and $\tilde K = \hat J$, we finally get the following equations of motion for the model having $SB(2,\mathbb{C})$ as target configuration space.
\beqa \label{eqmotdu}
\dot {\hat I}^k&=&2 i\tau a^2{\epsilon_p}^{k3}\partial_{\sigma}\hat{I}^p-2 a^2\tau^2 h^{pk}\partial_{\sigma} \hat{J}_p\nonumber\\
\dot {\hat J}_k&=& 2 i \tau a^2 {\epsilon_k}^{p3} \partial_{\sigma} \hat{J}_p.
\eeqa

To summarise, we were not able to recover the 'natural' Hamiltonian model  with Hamiltonian $H_1$ from the dual family obtained in Sec. \ref{twoparam},  but we managed to define another model with the same target phase space,  $T^*SB(2,\mathbb{C})$,  but different metric tensor, which can be  related  to  the dual family of Hamiltonians, \eqn{dualham} in the limit $i \alpha \rightarrow 0$ once a deformation has been performed. 
This establishes the wanted connection.

Finally, it is interesting to notice that it would have been impossible to obtain such a connection for the PCM where  the  WZ term is absent. Indeed,  for $a=0$ the Hamiltonian $H_2$ is identically zero  and  the equations of motion in (\ref{eqmotdu}) loose their significance. 

To conclude this section, let us shortly address the issue of  space-time symmetries.  Since the model is Lagrangian, the energy-momentum tensor may be obtained from the action \eqn{sbactionf}, yielding
\begin{equation}\label{enmomtensorsb}
T_{00}=T_{11}=\frac{1}{4\lambda^2}\left(\hat I^i h_{ij} \hat I^j + \hat J_i h^{ij} \hat J_j\right) ; \quad T_{01}=T_{10}=\frac{1}{4\lambda^2} \left(\hat I^i \delta_i^j \hat J_j\right),
\end{equation}
which is formally the same as the $SU(2)$ tensor \eqn{enmomtensor}. However,  our product is not bi-invariant,  that is, it doesn't satisfy 
\be
{f^{ab}}_{d} \, g^{cd}+{f^{ac}}_{d} \, g^{b d}=0,
\ee
which is a sufficient condition for Poincar\'e invariance (see for instance \cite{nappi}). Nonetheless, it is immediate to check that $T_{\mu\nu}$ is conserved and traceless. Moreover,  the Master Virasoro equations (\ref{virasoro}) are satisfied as well, so the model is conformally and Poincar\'e invariant at the classical level.

\section{Double WZW model}
\label{sectdoubledwzw}

So far we have been  able to give a description of the $SU(2)$ WZW model current algebra as the affine algebra of $SL(2,\mathbb{C})$ and to construct a map to a family of dual models, with the same current algebra and target phase space, but with momenta and configuration fields exchanged.
It is therefore natural to look for an action with manifest $SL(2,\mathbb{C})$ symmetry which could accommodate both models,  by doubling the number of degrees of freedom.  

Let us consider the $SL(2,\mathbb{C})$-valued field $\Phi: \Sigma \ni (t,\sigma) \rightarrow \gamma \in SL(2,\mathbb{C})$ and introduce the $\mathfrak{sl}(2,\mathbb{C})$-valued Maurer-Cartan one-forms $\gamma^{-1}d \gamma$. We postulate the following action functional with $SL(2,\mathbb{C})$ as target configuration space:
\begin{equation}
\mathcal{S}=\kappa_1 \int_{\Sigma} \left( \left( \Phi^{*}[\gamma^{-1}d\gamma] \stackrel{\wedge}{, } * \Phi^{*}[\gamma^{-1}d\gamma]  \right) \right)_N +\kappa_2 \int_{\mathcal{B}} \left( \left(\tilde{\Phi}^{*}\big[\tilde{\gamma}^{-1}d\tilde{\gamma} \stackrel{\wedge}{, } \tilde{\gamma}^{-1}d\tilde{\gamma} \wedge \tilde{\gamma}^{-1}d\tilde{\gamma} \big] \right) \right)_N
\end{equation}
with $\kappa_1, \kappa_2$  constants left arbitrary,  and the Hermitian product \eqn{newproduct} is employed. 

The equations of motion can be derived by following the same steps as in the previous section for the model on $SB(2,\mathbb{C})$, with the only difference that now the fields are valued in the Lie algebra  $\mathfrak{sl}(2,\mathbb{C})$  with structure constants ${C_{IJ}}^K$. We obtain 
\begin{equation}
\mathcal{H}_{(N)LK }\left(\partial_t A^K-\partial_{\sigma}J^K \right)-{C_{PL}}^Q \mathcal{H}_{ (N)QK}\left(A^P A^K-J^P J^K \right) =-2\frac{\kappa_2}{ \kappa_1} \, \mathcal{H}_{(N)QL } {C_{PS}}^Q A^P J^S,
\end{equation}
where we denoted by $A^I \equiv \left(A^i , L_i\right)$, $J^{I} \equiv \left(J^i, M_i \right)$ the $T SL(2,\mathbb{C})(\mathbb{R})$ coordinates, with double index notation and $\mathcal{H}_{(N)}$ is the Hermitean product defined in \eqn{newproduct}. 
The generalised doubled action so constructed describes a non-linear sigma model with Wess-Zumino term with target configuration space the group manifold of $SL(2,\mathbb{C})$.

\subsection{Doubled Hamiltonian description}

In order to describe the doubled model in the Hamiltonian formalism, let us start by considering only the kinetic term ($\kappa_2=0$). In this case the equations of motion can be written as
\begin{equation}
\mathcal{H}_{LK (N)}\left(\partial_t A^K-\partial_{\sigma}J^K \right)-{C_{PL}}^Q \mathcal{H}_{QK (N)}\left(A^P A^K-J^P J^K \right)=0,
\end{equation}
and we can define a genuine Lagrangian density as follows:
\begin{equation}
\mathscr{L}=\kappa_1 \mathcal{H}_{LK (N)}\left(A^L A^K-J^L J^K \right),
\end{equation}
from which canonical momenta can be defined
\begin{equation}
I_K \equiv \left(I_k, N^k \right)=\frac{\delta L}{\delta J^K}=2 \kappa_1 \mathcal{H}_{KL (N)} A^L,
\end{equation}
leading to the following Hamiltonian:
\begin{equation}\label{parentham}
H=\kappa_1 \int_{\mathbb{R}} d\sigma \left[ \left(\mathcal{H}^{-1}\right)^{LK (N)} I_L I_K+ \mathcal{H}_{LK (N)}J^L J^K  \right].
\end{equation}
The equal-time Poisson brackets can then be obtained in the usual way \cite{marottapcm}, resulting in 
\begin{equation}
\begin{aligned}\left\{I_{L} (\sigma), I_{K} (\sigma')\right\} &=\frac{1}{2\kappa_1} {C_{LK}}^{P} I_P \, \delta(\sigma-\sigma') \\\left\{I_{L}(\sigma), J^{K} (\sigma')\right\} &=\frac{1}{2\kappa_1} \left[ {C_{P L}}^{K} J^P \delta(\sigma-\sigma')-\delta_{L}^{K} \delta^{\prime}(\sigma-\sigma')\right] \\\left\{J^{L}(\sigma), J^K (\sigma')\right\} &=0, \end{aligned}
\end{equation}
and together with the Hamiltonian, they lead to the equations of motion
\begin{equation}
\partial_t I_K(\sigma)=\frac{1}{2\kappa_1}\left[\left(\mathcal{H}^{-1}\right)^{LP (N)}{C_{LK}}^Q I_Q(\sigma) I_P (\sigma)-\mathcal{H}_{LP (N)} \, {C_{QK}}^L J^Q(\sigma) J^P(\sigma)+\mathcal{H}_{KL (N)} \partial_{\sigma}J^L(\sigma) \right].
\end{equation}
Following the same approach we used for the $SB(2,\mathbb{C})$ model case, we can now include the WZ term, resulting in the modification of the equations of motion
\begin{equation}
\begin{aligned}
\partial_t I_K(\sigma) = {} &  \frac{1}{2\kappa_1}\Big[\left(\mathcal{H}^{-1}\right)^{LP (N)}{C_{LK}}^Q I_Q(\sigma) I_P (\sigma)-\mathcal{H}_{LP (N)} \, {C_{QK}}^L J^Q(\sigma) J^P(\sigma)+\mathcal{H}_{KL (N)} \partial_{\sigma}J^L(\sigma) \\ & -2\frac{\kappa_2}{ \kappa_1} \mathcal{H}_{QK (N)} \left(\mathcal{H}^{-1} \right)^{RP (N)} {C_{PS}}^Q  I_R J^S \Big],
\end{aligned}
\end{equation}
which can be obtained from the same Hamiltonian but modifying the Poisson structure as follows:
\begin{equation}
\begin{aligned}\left\{I_{L} (\sigma), I_{K} (\sigma')\right\} &=\frac{1}{2\kappa_1} {C_{LK}}^{P} I_P (\sigma) \, \delta(\sigma-\sigma')-\frac{\kappa_2}{\kappa_1^2}\, \mathcal{H}_{QK (N)} {C_{LP}}^Q J^P (\sigma) \delta(\sigma-\sigma') \\\left\{I_{L}(\sigma), J^{K} (\sigma')\right\} &=\frac{1}{2\kappa_1} \left[{C_{P L}}^{K} J^P (\sigma) \delta(\sigma-\sigma')-\delta_{L}^{K} \delta^{\prime}(\sigma-\sigma')\right] \\\left\{J^{L}(\sigma), J^K (\sigma')\right\} &=0. \end{aligned}
\end{equation}
Models with target configuration space $SU(2)$ or $SB(2,\C)$ could then be obtained by constraining the Hamiltonian \eqn{parentham}. The Lagrangian approach adopted in  \cite{marottapcm}, which requires to gauge one of the global symmetries of the parent action, presents some difficulties,  since minimal coupling is not enough anymore and there may be obstructions to be dealt  with. Indeed, although minimal coupling produces a gauge-invariant action, the equations of motion still depend on the extension to the $3$-manifold $\mathcal{B}$.
This issue is addressed e.g. in \cite{Hull:1990ms, Figueroa-OFarrill:1994uwr, Figueroa-OFarrill:1994vwl}. Besides that,  another problem, which is specific of the model, might affect the gauging.  In fact, in the cited references  the  gauged action  is always formulated for a semisimple group with  a Cartan-Killing metric. However, here in order to reproduce the $SB(2,\C)$ model we need to work with an Hermitian product. It is not clear how to handle the problem in this case and further investigation is needed, which shall be performed elsewhere.

\section{Conclusions and Outlook}
\label{sectconclusions}

This work further extends the analysis of the $SU(2)$ Principal Chiral Model performed by some of the authors in \cite{marottapcm}.

Starting from a canonical generalisation of the Hamiltonian picture associated to the WZW model with $SU(2)$ target configuration space, which consists in describing the dynamics of the model in terms of a one-parameter family of Hamiltonians and $SL(2,\C)$ Kac-Moody algebra of currents, we have highlighted the Drinfel'd double nature of the phase space, by introducing a further parameter both in the Hamiltonian and in the Poisson algebra. Our first result has been to show the Poisson-Lie symmetry of the model. Then, by performing a duality transformation in target phase space, we have been able to obtain a two-parameter  family of models which are Poisson-Lie dual to the previous ones by construction. The two families share the same target phase space, the group manifold of $SL(2,\C)$, but have configuration spaces  which are dual to each other, namely $SU(2)$ and its Poisson-Lie dual, $SB(2,\C)$. 
Although they have not been derived from an action principle, it has been shown that it is possible to exhibit an action, by means of an inverse Legendre transform which involves the symplectic form and the Hamiltonian.

As a natural step, we have  investigated the possibility of defining a Lagrangian WZW model with target tangent space $TSB(2,\C)$.  Being the group $SB(2,\C)$ not semi-simple, the problem of defining a non-degenerate product on its Lie algebra has been addressed, and a solution has been proposed. Once accomplished the Lagrangian picture, we have derived the Hamiltonian description on the cotangent space $T^*SB(2,\C)$. We have shown that, although its current algebra is obtained as the limit $\alpha\rightarrow 0$ of the $SL(2,\C)$ Kac-Moody algebra related to the dual family, it is not possible to obtain the Hamiltonian in the same limit, through a continuous deformation of phase spaces. It is however possible to define on $T^*SB(2,\C)$ a new Hamiltonian, in terms  of an alternative $O(3,3)$ metric. Such a model may be  related to  the dual family of $SL(2,\C)$ models  if one first performs  a deformation of the dynamics and then the limit $\alpha\rightarrow 0$. 
 It is interesting to notice  that such a connection relies on the presence of the WZ term, and the whole construction loses significance if the WZW coefficient is put to zero. A diagrammatic summary of the different models with corresponding relations between them is depicted in Fig. (\ref{fig:diagram}), where $Q$, $\Gamma$ and $\mathfrak{c}$ denote the target configuration space, phase space and current algebra respectively.
 
\begin{figure}[h!]
\begin{center}
\includegraphics[width=0.63\linewidth]{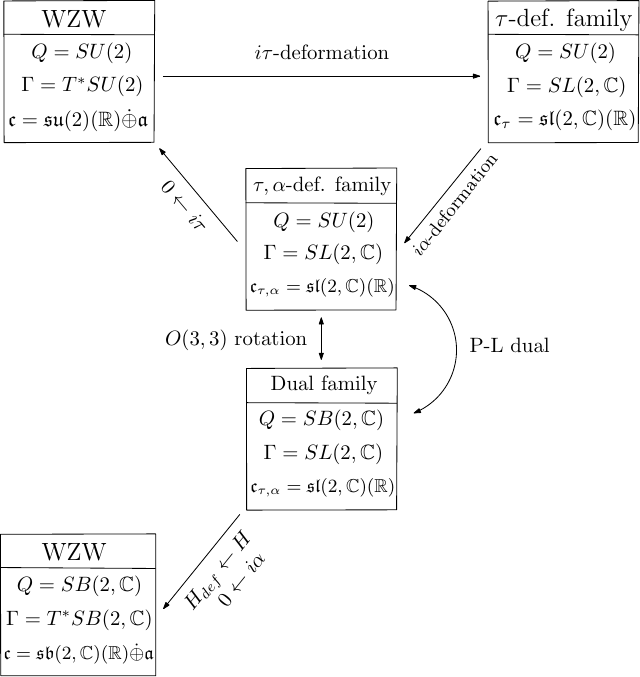}
\end{center}
\caption{Diagrammatic summary of the models considered and their relations. $Q$, $\Gamma$ and $\mathfrak{c}$ denote configuration space, phase space and current algebra respectively.}
\label{fig:diagram}
\end{figure}

Having introduced a well-defined WZW action on $SB(2,\mathbb{C})$ we have analysed the geometry of the target space as a string background solution. This is a non-compact  Riemannian hypersurface, whose metric is induced by a Lorentzian metric. The $\mathds{B}$-field and its flux have been calculated as well. 

Finally, we have addressed the possibility of making manifest  the $SL(2,\C)$ symmetry of both families of WZW models, by doubling the degrees of freedom and introducing a parent action with target configuration space  the Drinfel'd   double $SL(2,\mathbb{C})$. A doubled Hamiltonian formulation has been proposed,  such that a  restriction to either subgroup, $SU(2)$ or $SB(2,\C)$,  leads to the Hamiltonian formulation of the two sub-models.

As for future perspectives,  it would be interesting to quantise the interpolating model, and since it depends  on two further parameters, it would be worth looking at conformal invariance  in the quantum regime. In this respect however,   it should be recalled that finite dimensional irreducible representations of $\mathfrak{sl}(2,\mathbb{C})$ are non-unitary (see \cite{RSV93} for related analysis of the one-parameter family, in the case of $\tau$ real). On the other hand, such an alternative formulation seems to be well suited for a formal quantisation in the sense of Drinfel'd \cite{reshetikhin}, and this possibility could be  explored in the future.

As a further goal, we hope this work may contribute to the analysis of  string theories on $AdS$ geometries, the study of which would be interesting from the $AdS/CFT$ correspondence perspective. 

 \vspace{5pt}

\noindent{\bf Acknowledgements} 

We are grateful to Giuseppe Marmo and Vincenzo Emilio Marotta for useful discussions and suggestions. F. B. would like to thank Ivano Basile for stimulating discussions.

\begin{appendix}
\section{Poisson-Lie groups and Drinfel'd double structure of $SL(2,\mathbb{C})$}
\label{sl2c}

In this appendix we briefly review the mathematical setting of Poisson-Lie groups and Drinfel'd doubles, see \cite{semenov, alekseev:articolo, liu1997, charibook, collardtesi} for details. In particular we  focus on $SL(2,\mathbb{C})$ as a specific  example of Drinfel'd double since it  plays a major role throughout this paper.

\theoremstyle{definition}
\begin{definition}{} \label{defA1}
A \textit{Poisson-Lie group} is a Lie group $G$ with a Poisson structure such that the multiplication $\mu: G \times G \to G$ is a Poisson map if $G \times G$ is equipped with the product Poisson structure.
\end{definition}
Let $\mathfrak{g}$ denote its  Lie algebra, identified with $T_{e} G$, the tangent space at the group identity $e$. We can use the  Poisson brackets defined  on the group manifold to introduce  a Lie bracket on $\mathfrak{g}^*$, the dual vector space of $\mathfrak{g}$,  as follows:
\begin{definition}{}
The \textit{induced dual Lie bracket} on $\mathfrak{g}^*$, via the Poisson bracket $\{\cdot, \cdot\}$ on $G$ can be obtained as
\begin{equation}
[\xi_1, \xi_2 ]_{\mathfrak{g}^*}=d \{f_1, f_2 \}(e),
\end{equation}
with $f_1, f_2 \in C^{\infty}(G)$ with the property $df_1(e)=\xi_1$, $df_2(e)=\xi_2$.
\end{definition}
It is possible to prove that this induced bracket is indeed a Lie bracket. The compatibility condition between Lie and Poisson structures gives the following relation
\begin{equation}
\label{compcondpl}
\left\langle[X, Y],[u, v]^{*}\right\rangle+\left\langle\mathrm{ad}_{v}^{*} X, \mathrm{ad}_{Y}^{*} v\right\rangle-\left\langle\mathrm{ad}_{v}^{*} X, \mathrm{ad}_{Y}^{*} u\right\rangle-\left\langle\mathrm{ad}_{u}^{*} Y, \mathrm{ad}_{X}^{*} v\right\rangle+\left\langle\mathrm{ad}_{v}^{*} Y, \mathrm{ad}_{X}^{*} u\right\rangle=0,
\end{equation}
with $u, v \in \mathfrak{g}^*$ and $X, Y \in \mathfrak{g}$, while $\text{ad}^*_X$ and $\text{ad}^*_u$ respectively denote the coadjoint actions of $\mathfrak{g}$ and $\mathfrak{g}^*$ on each other and $\langle \cdot, \cdot \rangle$ is the symmetric pairing on $\mathfrak{g}$. This construction allows to define a Lie bracket on the direct sum $\mathfrak{g} \oplus \mathfrak{g}^*$ as follows:
\begin{equation}\label{dbracket}
[X+\xi, Y+\zeta]=[X, Y]+[\xi, \zeta]^{*}-a d_{X}^{*} \zeta+a d_{Y}^{*} \xi+a d_{\zeta}^{*} X-a d_{\xi}^{*} Y,
\end{equation}
with $X, Y \in \mathfrak{g}$ and $\xi, \zeta \in \mathfrak{g}^*$.

A Lie algebra with a compatible dual Lie bracket is called a \textit{Lie bialgebra}. If the group $G$ is connected, the compatibility condition is enough to integrate $[\cdot,\cdot]^*$ to a Poisson structure on it, making it Poisson-Lie, and the Poisson structure is unique. Since the role of $\mathfrak{g}$ and $\mathfrak{g}^*$ in (\ref{compcondpl}) is symmetric, one has also a Poisson-Lie group $G^*$ with Lie algebra $(\mathfrak{g}^* ,[ \cdot , \cdot ]^*)$ and a Poisson structure whose linearization at $e \in G^*$ gives the bracket $[\cdot , \cdot ]$. In this case $G^*$ is said to be the Poisson-Lie dual group of $G$.

The triple $(\mathfrak{d}, \mathfrak{g}, \mathfrak{g}^*)$ where $\mathfrak{d} = \mathfrak{g} \oplus \mathfrak{g}^*$ is a Lie algebra with bracket given by \eqn{dbracket} is known as a Manin triple, whereas its exponentiation to a Lie group $D$ is the Drinfel'd double of $G$.  More precisely
\vspace{0.5pt}
\begin{definition}{}
A \textit{Drinfel'd double} is an even-dimensional Lie group $D$ whose Lie algebra $\mathfrak{d}$ can be decomposed into a pair of maximally isotropic subalgebras\footnote{An \textit{isotropic subspace} of $\mathfrak{d}$ with respect to $\langle \cdot, \cdot \rangle$ is defined as a subspace $A$ on which the bilinear form vanishes: $\langle a, b \rangle=0 \, \, \, \forall \, a, b \in A$. An isotropic subspace is said to be \textit{maximal} if it cannot be enlarged while preserving the isotropy property, or, equivalently, if it is not a proper subspace of another isotropic space.}, $\mathfrak{g}$ and $\tilde{\mathfrak{g}}$, with respect to a non-degenerate (ad)-invariant bilinear form $\langle \cdot, \cdot \rangle$ on $\mathfrak{d}$.
\end{definition}
\begin{definition}{}
A \textit{Manin triple} $(\mathfrak{c}, \mathfrak{a}, \mathfrak{b})$ is a Lie algebra with a non-degenerate scalar product $\langle \cdot, \cdot \rangle$ on $\mathfrak{c}$ such that:

(i) $\langle \cdot, \cdot \rangle$ is invariant under the Lie bracket: $\langle c_1, [c_2, c_3] \rangle=\langle [c_1, c_2], c_3 \rangle, \quad \forall c_1,c_2,c_3 \in \mathfrak{c}$;

(ii) $\mathfrak{a}, \mathfrak{b}$ are maximally isotropic Lie subalgebras with respect to $\langle \cdot, \cdot \rangle$;

(iii) $\mathfrak{a}, \mathfrak{b}$ are complementary (as linear subspaces), i.e. $\mathfrak{c}=\mathfrak{a} \oplus \mathfrak{b}$.
\end{definition}

Note that since the bilinear form is non-degenerate by definition, we can identify $\tilde{\mathfrak{g}}$ with the dual vector space $\mathfrak{g}^*$, and the Lie subalgebra structure on $\tilde{\mathfrak{g}}$ then makes $\mathfrak{d}$ into a Lie bialgebra. It is possible to prove that,  conversely, every Lie bialgebra defines a Manin triple by identifying $\tilde{\mathfrak{g}}=\mathfrak{g}^*$ and defining the mixed Lie bracket between elements of $\mathfrak{g}$ and $\tilde{\mathfrak{g}}$ in such a way to make the bilinear form invariant. Indeed, one can prove that if we want to make $\mathfrak{d}=\mathfrak{g}\oplus \tilde{\mathfrak{g}}$ into a Manin triple, using the natural scalar product on $\mathfrak{d}$, there is only one possibility for the Lie bracket, as explained in the following.

\begin{lemma}
Let $\mathfrak{g}$ be a Lie algebra with Lie bracket $[\cdot, \cdot]$ and dual Lie bracket $[\cdot, \cdot]_{\mathfrak{g}^*}$. Every Lie bracket on $\mathfrak{d}=\mathfrak{g} \oplus \mathfrak{g}^*$ such that the natural scalar product is invariant and such that $\mathfrak{g}$, $\mathfrak{g}^*$ are Lie subalgebras is given by:
\begin{equation}
\label{drinfbrack}
\begin{aligned}
&[x, y]_{\mathfrak{d}}=[x, y] \quad \quad \quad \quad \quad \quad \, \, \, \,  \forall \, x, y \in \mathfrak{g} \\
&[\alpha, \beta]_{\mathfrak{d}}=[\alpha, \beta]_{\mathfrak{g}^{*}} \quad \quad \quad \quad \quad \, \, \, \, \forall \, \alpha, \beta \in \mathfrak{g}^* \\
&[x, \alpha]_{\mathfrak{d}}=-\operatorname{ad}_{\alpha}^{*} x+\operatorname{ad}_{x}^{*} \alpha \quad \quad \forall \, x \in \mathfrak{g}, \alpha \in \mathfrak{g}^*.
\end{aligned}
\end{equation}
\end{lemma}
In order for the whole algebra to satisfy  Jacobi identity the brackets on the two dual spaces have to be compatible. Moreover,  this bracket is the unique Lie bracket which makes $(\mathfrak{d}, \mathfrak{g}, \mathfrak{g}^*)$ into a Manin triple.

To make things more explicit, on choosing $T_i$ and $\widetilde{T}^i$ as the generators of the Lie algebras $\mathfrak{g}$ and $\tilde{\mathfrak{g}}$ respectively, such that $T_I \equiv (T_i, \widetilde{T}^i)$ are the generators of $\mathfrak{d}$, by the property of isotropy and duality as vector spaces we have
\begin{equation}
\begin{aligned}
\langle T_i, T_j\rangle &= 0 \\
\langle\widetilde{T}^i, \widetilde{T}^j\rangle &= 0 \\
\langle T_i, \widetilde{T}^j\rangle &={\delta_i}^j,
\end{aligned}
\end{equation}
with $i=1,2, \dots, \text{dim}\, G$, while the bracket in (\ref{drinfbrack}), in doubled notation given by $[T_I, T_J]={F_{IJ}}^K T_K$, can be written explicitly as follows:
\begin{equation}
\label{drinfbrackexpl}
\begin{aligned}
&[T_{i}, T_{j}]={f_{i j}}^k T_k \\ &
[\widetilde{T}^{i}, \widetilde{T}^{j}]={g^{i j}}_k \widetilde{T}^{k} \\ &
[T_{i}, \widetilde{T}^{j}]={f_{k i}}^j \widetilde{T}^{k}-g^{{k j}}_i T_{k},
\end{aligned}
\end{equation}
with ${f_{i j}}^k$, ${g^{i j}}_k$, ${F_{IJ}}^K$ structure constants for $\mathfrak{g}$, $\tilde{\mathfrak{g}}$ and $\mathfrak{d}$ respectively.

 Jacobi identity on $\mathfrak{d}$, or equivalently the compatibility condition, impose the following  constraint on   structure constants of  dual algebras $\mathfrak{g}$ and $\tilde{\mathfrak{g}}$:
\begin{equation}
\label{bialgebracond}
{g^{p k}}_i {f_{q p}}^j-{g^{p j}}_i f_{q p}^k-{g^{p k}}_q f_{i p}^j+{g^{p j}}_q f_{i p}^k-{g^{j k}}_p f_{q i}^p=0,
\end{equation}
which is equivalent to Eq. \eqn{compcondpl}, obtained as a compatibility condition between Poisson and group structure on a given  group G (Poisson-Lie condition). From previous results  some observations follow: the relation is completely symmetric in the structure constants of the dual partners as the entire construction is symmetric, and exchanging the role of the two subalgebras leads exactly to the same structure. This will be important for the formulation of  Poisson-Lie duality. It is worth to note that this condition is always satisfied whenever at least one of the two subalgebras is Abelian. This means that if $\mathfrak{d}$ is a Lie algebra of dimension $2d$, we always have at least two Manin triples $(\mathfrak{g},\mathbb{R}^d )$ and $(\mathbb{R}^d ,\mathfrak{g})$, with $\text{dim} \, \mathfrak{g}=d$.

By exponentiation of $\mathfrak{g}$ and $\tilde{\mathfrak{g}}$ one gets the dual Poisson-Lie groups $G$ and $\tilde{G}$ such that, in a given local parametrization, $D= G \cdot \tilde{G}$, or by changing parametrization, $D= \tilde G \cdot {G}$. 
The simplest example is the cotangent bundle of any $d$-dimensional Lie group $G$, $T^* G \simeq G \ltimes \mathbb{R}^d$, which we shall call the classical double, with trivial Lie bracket for the dual algebra $\tilde{\mathfrak{g}} \simeq \mathbb{R}^d$. 

The natural symplectic structure on the group manifold of the double $D$ is the so called Semenov-Tian-Shansky structure \cite{semenovtianshansky85} $\{f, g \}_D$, for $f, g$ functions on $D$. If one considers the functions $f, g$ to be invariant with respect to the action of the group $\tilde{G}$ ($G$) on $D$, they can be basically interpreted as functions on the group manifold of $G$ ($\tilde{G}$), which then inherit the Poisson structure directly from the double.

We finally point out that there may be many decompositions of $\mathfrak{d}$ into maximally isotropic subspaces, which are not necessarily subalgebras: when the whole mathematical setting is applied to sigma models, the set of all such decompositions  plays the role of the modular space of sigma models mutually connected by a $O(d,d)$ transformations. In particular, for the manifest Abelian T-duality of the string model on the d-torus, the Drinfel'd double is $D = U(1)^{2d}$ and its modular space is in one-to-one correspondence with $O(d,d;\mathbb{Z})$ \cite{klimcik3}.

After this brief review of Drinfel'd doubles and Manin triples, for the purposes of this work we will focus on a particular example of Drinfel'd double, $SL(2,\mathbb{C})$.

As a starting point let us fix the notation. The  real $\mathfrak{sl}(2,\mathbb{C})$  Lie algebra is usually represented in the form:
\begin{equation}
\label{slalgebra}
\begin{aligned}
& \left[e_{i}, e_{j}\right]=i {\epsilon_{i j}}^{k} e_{k} \\
&\left[b_{i}, b_{j}\right]=-i {\epsilon_{i j}}^{k} e_{k} \\
&\left[e_{i}, b_{j}\right]=i {\epsilon_{i j}}^{k} b_{k},
\end{aligned}
\end{equation}
with $\{e_i\}_{i=1,2,3}$  generators of  the $\mathfrak{su}(2)$ subalgebra, $\{b_i\}_{i=1,2,3}$ boosts generators. The linear combinations  
\begin{equation} \label{sbgen}
\hat{e}^{i}=\delta^{i j}\left(b_{j}+{\epsilon^k}_{j 3} e_{k}\right),
\end{equation}
are  dual to the ${e_i}$ generators with respect to the Cartan-Killing product naturally defined on $\mathfrak{sl}(2,\mathbb{C})$ as $\langle v,w \rangle=2\left[\text{Im}\left(vw \right) \right], \, \,  \forall \, v,w \in \mathfrak{sl}(2,\mathbb{C})$. Indeed, it is easy to show that
\begin{equation}
\left\langle\hat{e}^{i}, e_{j}\right\rangle= 2 \operatorname{Im}\left[\operatorname{Tr}\left(\hat{e}^{i} e_{j}\right)\right]=\delta_{j}^{i}.
\end{equation}
Moreover, the dual vector space $\mathfrak{su}(2)^*$ spanned by $\{\hat{e}^i\}_{i=1,2,3}$ is  the Lie algebra of the Borel subgroup of $SL(2,\mathbb{C})$, so called $SB(2,\mathbb{C})$,  of $2 \times 2$ upper triangular complex-valued matrices with unit determinant and real diagonal, for which the Lie bracket is defined as follows
\begin{equation}
\label{sbalg}
\left[\hat{e}^{i}, \hat{e}^{j}\right]=i {f^{i j}}_{k} \hat{e}^{k}
\end{equation}
and
\begin{equation}
\label{mixedsusb}
\left[\hat{e}^{i}, e_{j}\right]=i \epsilon_{j k}^{i} \hat{e}^{k}+i e_{k} f^{k i},
\end{equation}
with structure constants ${f^{ij}}_k= \epsilon^{ijs}\epsilon_{s3k}$.
As a manifold $SB(2,\mathbb{C})$ is non-compact and its Lie algebra is non-semisimple, which is reflected in the fact that the structure constants ${f^{ij}}_k$ as previously defined are not completely antisymmetric.

It is important to note that  the following relations hold
\begin{equation}
\left\langle e_{i}, e_{j}\right\rangle=\left\langle\hat{e}^{i}, \hat{e}^{j}\right\rangle= 0,
\end{equation}
so that both subalgebras $\mathfrak{su}(2)$ and $\mathfrak{sb}(2,\mathbb{C})$ are maximal isotropic subalgebras of $\mathfrak{sl}(2,\mathbb{C})$ with respect to $\langle \cdot, \cdot \rangle$. Therefore, $(\mathfrak{sl}(2,\mathbb{C}), \mathfrak{su}(2), \mathfrak{sb}(2,\mathbb{C}))$ is a Manin triple with respect to the natural Cartan-Killing pairing on $\mathfrak{sl}(2,\mathbb{C})$ and $SL(2,\mathbb{C})$ is a Drinfel'd double with respect to this decomposition (polarization): $SL(2,\mathbb{C})= SU(2) \cdot SB(2,\mathbb{C})$.

Let us observe that the first of the Lie brackets  (\ref{slalgebra}) together with (\ref{sbalg}) and (\ref{mixedsusb}) have exactly the form (\ref{drinfbrackexpl}) and that in doubled notation, $e_I=\begin{pmatrix}
  e_i  \\
  \hat{e}^i
 \end{pmatrix},$ with $e_i \in \mathfrak{su}(2)$ and $\hat{e}^i \in \mathfrak{sb}(2,\mathbb{C})$, the scalar product 
\begin{equation}
\label{o33metric}
\left\langle e_I, e_J\right\rangle=\eta_{I J}=\left(\begin{array}{cc}
{0} & {{\delta_{i}}^{j}} \\
{{\delta^{i}}_{j}} & {0}
\end{array}\right)
\end{equation}
corresponds to an $O(3,3)$ invariant metric.

Other than the natural Cartan-Killing bilinear form there is also another non-degenerate invariant scalar product which can be defined on $\mathfrak{sl}(2,\mathbb{C})$ as:
\begin{equation}
(v,w)=2 \text{Re}\left[\text{Tr}\left(vw \right) \right], \quad \forall \, v,w \in \mathfrak{sl}(2,\mathbb{C}).
\end{equation}
However, it is easy to check that $\mathfrak{su}(2)$ and $\mathfrak{sb}(2,\mathbb{C})$ are no longer isotropic subspaces with respect to this scalar product, it being
\begin{equation}
(e_i, e_j)=\delta_{ij}, \quad (b_i, b_j)=-\delta_{ij}, \quad (e_i, b_j)=0.
\end{equation}
Note that this does not give rise to a positive-definite metric. However, on denoting by $C_+$, $C_-$ respectively the two subspaces spanned by $\{e_i \}$ and $\{ b_i\}$, the splitting $\mathfrak{sl}(2,\mathbb{C})=C_+ \oplus C_-$ (which is not a Manin triple polarization by the way, since $C_+$ and $C_-$ do not close as subalgebras) defines a positive definite metric $\mathcal{H}$ on $\mathfrak{sl}(2,\mathbb{C})$ as follows:
\begin{equation}
\mathcal{H}=\left( , \right)_{C_+}-\left( , \right)_{C_-}.
\end{equation}
This is a Riemannian metric and we denote it with the symbol $\left( \left(	\, ,\,  \right) \right)$. In particular:
\begin{equation}
\left( \left(e_i, e_j \right) \right) \equiv  \left(e_i, e_j \right), \quad \left( \left( b_i, b_j \right) \right) \equiv -\left(b_i, b_j \right), \quad \left( \left(e_i, b_j \right) \right) \equiv \left(e_i ,b_j\right)=0.
\end{equation}
In doubled notation, $e_I=\begin{pmatrix}
  e_i  \\
  \hat{e}^i
 \end{pmatrix}$, this Riemannian product can be written instead as
\begin{equation}
\label{riemannianmetric}
\left( \left( e_I , e_J\right) \right)=\mathcal{H}_{IJ}=\begin{pmatrix}
  \delta_{ij} & -\delta_{ip}\epsilon^{jp3} \\
  -\epsilon^{i p3} \delta_{pj} & \delta^{ij}+\epsilon^{il3}\delta_{\ell k} \epsilon^{j k3}
 \end{pmatrix},
\end{equation}
which satisfies the relation $\mathcal{H}\eta \mathcal{H}=\eta$, indicating that $\mathcal{H}$ is a pseudo-orthogonal $O(3,3)$ matrix. 

This product can be verified to be equivalent to
\begin{equation}
\label{sb2cproduct}
((u,v)) \equiv 2 \text{Re}\left[\text{Tr}\left( u^{\dagger} v \right) \right],
\end{equation}
and its restriction to the  $SB(2,\mathbb{C})$ subalgebra, which  will be indicated by $h$,  has the following form:
\begin{equation}
\label{sbmetric}
h^{ij}=\delta^{ij}+\epsilon^{i \ell 3}\delta_{\ell k}\epsilon^{jk3}.
\end{equation}

It is interesting to notice  that  the $O(3,3)$ metric in (\ref{o33metric}) and the pseudo-orthogonal metric  in (\ref{riemannianmetric}) respectively have the same structure as  the $O(d,d)$ invariant metric and the so called generalised metric $\mathcal{H}$ of Double Field Theory \cite{tseytlin1, tseytlin2, hullzw}. 

Finally, let us notice that the most general action functional involving fields valued in the Lie algebra $\mathfrak{sl}(2,\C)$ should contain a combination of the two products, \eqn{o33metric}  and \eqn{riemannianmetric}. This essentially amounts to consider the Hermitian product 
\begin{equation}
\label{newproduct}
\mathcal{H}_N=((u,v))_N \equiv \text{Tr}(u^{\dagger}v)
\end{equation}
which is indeed necessary   to define a non-vanishing WZW term for the $SB(2,\C)$ related model (see Sec. \ref{sectwzwsb2c}). 
When restricting to the  $\mathfrak{sb}(2,\mathbb{C})$ subalgebra it acquires the form
\begin{equation}
h_N^{ij}=\begin{pmatrix}
  1 & -i & 0\\
  i & 1 & 0\\
  0 & 0 & 1/2
 \end{pmatrix},
\end{equation}
and obviously satisfies the relation $h_N^{ij}+h_N^{ji}=h^{ij}$.  It is possible to check (see Sec. \ref{sectwzwsb2c}) that only its real, diagonal  part, namely \eqn{sb2cproduct}, contributes  when limited  to the quadratic term of the action \eqn{sbaction}, while only its imaginary, off-diagonal part, namely \eqn{o33metric}, contributes when computing the WZ term.  

\section{Symplectic form for the $SL(2,\mathbb{C})$ current algebra}\label{sympl}

In this section we will briefly sketch the derivation  of the symplectic form \eqn{symplecticformgl} for the two-parameter family of models 
obtained in Sec. \ref{sectduality}. 

The Poisson algebra we start with is the one in (\ref{dcalgebra}), which we report  for convenience:
\beqa \label{pbracketappendixb}
 \{\tilde{K}_i(\sigma), \tilde{K}_j(\sigma') \}&=&i\alpha{\epsilon_{ij}}^k \tilde{K}_k(\sigma)\delta(\sigma-\sigma')- \alpha^2 \hat C \delta_{ij}\delta'(\sigma-\sigma') \nonumber\\
\{\tilde{S}^i(\sigma), \tilde{S}^j(\sigma')\} &=&i\tau{f^{ij}}_k \tilde{S}^k(\sigma)\delta(\sigma-\sigma') +\tau^2 \hat C h^{ij} \delta'(\sigma-\sigma') \\ 
\{\tilde{K}_i(\sigma), \tilde{S}^j(\sigma')\}&=&\left[i\alpha{\epsilon_{ki}}^j \tilde{S}^k(\sigma)  +i\tau {f^{jk}}_i \tilde{K}_k(\sigma)  \right] \delta(\sigma-\sigma') +  ( i \alpha\hat C'\delta_i^j - i \tau \hat C {\epsilon_i}^{j3})  \delta'(\sigma-\sigma').\nonumber
\eeqa

Let, $X_{\tilde{K}}$ and $X_{\tilde{S}}$ indicate the  Hamiltonian vector fields associated with  the  currents, so  that $\omega(X_{\tilde{K}_i}, X_{\tilde{K}_j})=\{\tilde{K}_i(x), \tilde{K}_j(x)\}$,  with analogous expressions for the other brackets. They are left-invariant because so are the currents. On introducing  their dual one-forms $\theta^i, \hat\theta_i$, with   $\theta^i X_{\tilde{K}_j}={\delta^i}_j$ and $\hat{\theta}_i X_{\tilde{S}^j}={\delta_i}^j$,  the Poisson brackets in (\ref{pbracketappendixb}) can be easily obtained from the following symplectic form
\begin{equation}\label{symplformtheta}
\begin{aligned}
& {} \omega= \int_{\mathbb{R}^2} d\sigma \, d\sigma'   \bigg\{ \theta^i(\sigma) \wedge \theta^j(\sigma') \left[i\alpha{\epsilon_{ij}}^k \tilde{K}_k(\sigma)\delta(\sigma-\sigma')- \alpha^2 \hat C \delta_{ij}\delta'(\sigma-\sigma')\right]\\ &+ \hat{\theta}_i(\sigma) \wedge \hat{\theta}_j(\sigma') \left[i\tau{f^{ij}}_k \tilde{S}^k(\sigma)\delta(\sigma-\sigma') +\tau^2 \hat C h^{ij} \delta'(\sigma-\sigma') 	\right] \\ & + \theta^i(\sigma) \wedge \hat{\theta}_j(\sigma') 	\left[\left(i\alpha{\epsilon_{ki}}^j \tilde{S}^k(\sigma)  +i\tau {f^{jk}}_i \tilde{K}_k(\sigma) \right)\delta(\sigma-\sigma')+\left(i \alpha\hat C'\delta_i^j - i \tau \hat C {\epsilon_i}^{j3} \right) \delta'(\sigma-\sigma') \right] \bigg\}.
\end{aligned}
\end{equation}
The latter may be further manipulated and expressed in terms of  the original group valued fields $g \in SU(2)$ and $\ell \in SB(2,\mathbb{C})$. This can be obtained by means of the  left invariant Maurer-Cartan $1$-forms relative to  each of the two groups, which read explicitly
\begin{equation}
g^{-1}dg=i \theta^i e_i, \quad \ell^{-1}d\ell=i \hat{\theta}_i \hat{e}^i.
\end{equation}
Hence,  by defining 
\begin{equation}
-i\alpha \hat{C}g^{-1}\partial_{\sigma} g=i\delta^{kp}\tilde{K}_p e_k, \quad i\tau\hat{C} \ell^{-1}\partial_{\sigma} \ell=i(h^{-1})_{kp} \tilde{S}^p \hat{e}^k
\end{equation}
it is possible to show that the symplectic form  (\ref{symplformtheta}) can be written in terms of $g$ and $\ell$ as in (\ref{symplecticformgl}). Since it is not immediate to see that the two expressions are equal, we shall go through the main steps for the first term of (\ref{symplecticformgl}), namely $\int d\sigma \, \text{Tr}_{\mathcal{H}}\left[g^{-1}dg \wedge \partial_{\sigma}(g^{-1}dg) \right]$,   as for the others it works in the same way. The starting point is to decompose the Maurer-Cartan $1$-form in its Lie algebra components:
\begin{equation*}
\begin{aligned}
{} & \alpha^2 \hat{C}\int_{\mathbb{R}} d\sigma \, \text{Tr}_{\mathcal{H}}\left[g^{-1}dg \wedge \partial_{\sigma}(g^{-1}dg) \right] \\ & =\alpha^2 \hat{C}\int_{\mathbb{R}} d\sigma \, \text{Tr}_{\mathcal{H}}\left[-g^{-1}dg \wedge g^{-1}\partial_{\sigma}g g^{-1}dg+g^{-1}dg \wedge g^{-1}\partial_{\sigma}dg \right] \\ &
=-\alpha\int_{\mathbb{R}} d\sigma \text{Tr}_{\mathcal{H}}\left[\theta^i \wedge  \delta^{kp}\tilde{K}_p e_i e_k \theta^j e_j \right] \\ & +\alpha^2 \hat{C}\int_{\mathbb{R}^2} d\sigma d\sigma' \delta(\sigma-\sigma') \text{Tr}_{\mathcal{H}}\left[ g^{-1}dg(\sigma')\wedge g^{-1}\partial_{\sigma}dg(\sigma')\right] \\ & =i\alpha\int_{\mathbb{R}^2} d\sigma \, d\sigma'  \delta(\sigma-\sigma') \theta^i(\sigma) \wedge \theta^j(\sigma') {\epsilon_{ij}}^k \tilde{K}_k(\sigma) \\ & +\alpha^2 \hat{C} \int_{\mathbb{R}^2} d\sigma d\sigma' \partial_{\sigma}\left\{\delta(\sigma-\sigma') \text{Tr}_{\mathcal{H}}\left[ g^{-1}dg(\sigma')\wedge g^{-1}dg(\sigma)\right]\right\} \\ & +\alpha^2 \hat{C}\int_{\mathbb{R}^2} d\sigma d\sigma' \delta'(\sigma-\sigma')\text{Tr}_{\mathcal{H}}\left[ g^{-1}dg(\sigma)\wedge g^{-1}dg(\sigma')\right] \\ & =\int_{\mathbb{R}^2} d\sigma \, d\sigma'   \theta^i(\sigma) \wedge \theta^j(\sigma') \left[i\alpha{\epsilon_{ij}}^k \tilde{K}_k(\sigma)\delta(\sigma-\sigma')- \alpha^2 \hat C \delta_{ij}\delta'(\sigma-\sigma')\right],
\end{aligned}
\end{equation*}
which is indeed the first term in (\ref{symplformtheta}). In the last equation we used the antisymmetry property of the wedge product. Similar calculations can be performed to obtain the remaining terms.



\end{appendix}

\end{document}